\documentclass[12pt]{article}
\usepackage[utf8]{inputenc}
\usepackage[margin=2cm]{geometry}
\usepackage{epsfig,amsfonts,amssymb,framed,amsmath,xcolor}
\usepackage{slashed}
\usepackage{hyperref}
\usepackage{mathtools}
\usepackage{cancel}
\input epsf.sty
\usepackage{hhline,verbatim}
\usepackage{bm}
\usepackage{array,upgreek}
\usepackage{ytableau}

\usepackage{tikz, pgf}
\usetikzlibrary{shapes.misc}
\usetikzlibrary{snakes}
\usetikzlibrary{decorations.pathmorphing}
\usetikzlibrary{shapes}
\usetikzlibrary{arrows.meta}

\usetikzlibrary{fit}

\numberwithin{equation}{section}
\newcommand{\be}{\begin{eqnarray}}
\newcommand{\ee}{\end{eqnarray}}

\newcommand{\bi}{\begin{itemize}}
\newcommand{\ei}{\end{itemize}}

\newcommand{\bse}{\begin{subequations}}
\newcommand{\ese}{\end{subequations}}

\newcommand{\f}{\frac}

\newcommand{\p}{\partial}

\newcommand{\non}{\nonumber}

\newcolumntype{P}[1]{>{\centering\arraybackslash}p{#1}}

\definecolor{ablue}{rgb}{0.36, 0.54, 0.66}
\definecolor{afuchsia}{rgb}{0.57, 0.36, 0.51}
\definecolor{ao}{rgb}{0.0, 0.5, 0.0}
\definecolor{armygreen}{rgb}{0.29, 0.33, 0.13}
\definecolor{arsenic}{rgb}{0.23, 0.27, 0.29}
\definecolor{auburn}{rgb}{0.43, 0.21, 0.1}
\definecolor{asaurus}{rgb}{0.43, 0.5, 0.5}
\definecolor{bgrey}{rgb}{0.52, 0.52, 0.51}
\definecolor{bole}{rgb}{0.47, 0.27, 0.23}
\definecolor{brgreen}{rgb}{0.0, 0.26, 0.15}
\definecolor{burgundy}{rgb}{0.5, 0.0, 0.13}
\definecolor{byzantium}{rgb}{0.44, 0.16, 0.39}

\begin{document}



\baselineskip 24pt

\begin{center}
{\Large \bf  Revisiting Higher-Spin Gyromagnetic Couplings}

\end{center}

\vskip .6cm
\medskip

\vspace*{4.0ex}

\baselineskip=18pt

\begin{center}

{\large 
\rm Raffaele Marotta$^a$, Massimo Taronna$^{a,b}$ and Mritunjay Verma$^{c}$ }

\end{center}

\vspace*{4.0ex}

\centerline{ \it \small $^a$Istituto Nazionale di Fisica Nucleare (INFN), Sezione di Napoli, }
\centerline{ \it \small Complesso Universitario di Monte S. Angelo ed. 6, via Cintia, 80126, Napoli, Italy.}
\centerline{\it \small $^b$Dipartimento di Fisica ``Ettore Pancini'', Universita' degli Studi di Napoli,}
\centerline{\it \small  Federico II, Monte S. Angelo, Via Cintia, 80126 Napoli, Italy}

\centerline{\it \small $^c$Mathematical Sciences, Highfield, University of Southampton, SO17 1BJ Southampton, UK}

\vspace*{1.0ex}
\centerline{\small E-mail:  raffaele.marotta@na.infn.it, massimo.taronna@unina.it,
m.verma@soton.ac.uk }

\vspace*{5.0ex}

\centerline{\bf Abstract} \bigskip

We analyze Bosonic, Heterotic, and Type II string theories compactified on a generic torus having constant moduli. By computing the hamiltonian giving the interaction between massive string excitations and $U(1)$ gauge fields arising from the graviton and Kalb-Ramond field upon compactification, we derive a general formula for such couplings that turns out to be universal in all these theories. We also confirm our result by explicitly evaluating the relevant string three-point amplitudes. From this expression, we determine the gyromagnetic ratio $g$ of massive string states coupled to both gauge-fields. For a generic mixed symmetry state, there is one gyromagnetic coupling associated with each row of the corresponding Young Tableau diagram. For all the states having zero Kaluza Klein or Winding charges, the value of $g$ turns out to be $1$. We also explicitly consider totally symmetric and mixed symmetry states (having two rows in the Young diagram) associated with the first Regge-trajectory and obtain their corresponding $g$ value.

\vfill

\vfill \eject

\baselineskip18pt

\tableofcontents



\section{Introduction}

Charting the landscape of gravitational theories is a problem of key importance both to achieve a better understanding of string theory as well as to clarify the basic consistency requirements that gravitational theories should abide to. While the attention is often restricted to massless fields of spin up to two, a generic feature of consistent gravitational theories is the presence of infinite towers of massive fields of increasing spin.

In this work we focus on the leading electromagnetic coupling of massive excitations which is the first non-minimal correction to the minimal coupling. The problem of understanding how to couple consistently such massive excitations has a long history starting from Fierz and Pauli \cite{FierzPauli:1939} who were the first to analyse consistent electromagnetic interactions of massive spinning fields. The problem turned out to be more subtle than expected since just introducing a minimal coupling was not a consistent solution. It was quickly realised that minimal electromagnetic interactions must be supplemented by a non-minimal coupling proportional to the Electromagnetic tensor $F_{\mu\nu}$ and whose overall coupling constant is usually expressed in terms of the electric charge $q$ up to a numerical factor usually referred to as gyromagnetic ratio $g$. The gyromagnetic ratio determines the intensity of their magnetic moments in the interaction with external magnetic fields.

In the `50s, using the minimal coupling, Belinfante hypothesized that $g$ had to be equal to the inverse of the spin of the particle \cite{Belifante:1950}. This relation gives the correct value for a Dirac particle but it fails to give the correct tree level gyromagnetic ratio of the the other higher spin (HS) particles such as W-boson which turns out to be $g=2$. This is expected since, as mentioned above, the minimal coupling does not give rise to a consistent electromagnetic interaction of the higher spin fields.

The value $g=2$ naturally arises also from a number of different approaches which include the study of the precession of a spin particle in a magnetic field  \cite{BargmannMichelTelegdi:1959}, the high energy behaviour of scattering amplitudes \cite{Weinberg:1970} and string theory \cite{Ademollo:1974,ArgyresNappi:1989, Ferrara&Porrati&Telegdi,9809142} that suggest that the natural value for this ratio is $g=2$ (see also Ref. \cite{Holstein} for a review on the subject). 
Following these results, in the 90s, Ferrara, Porrati and Telegdi proposed an electromagnetic coupling which is consistent with the value $g=2$ for all the elementary particles of arbitrary spin \cite{Ferrara&Porrati&Telegdi}. The value $g=2$ has also been recently clarified in the context of massive HS interactions \cite{1204.1064,Cortese:2013lda,Rahman:2016tqc}. It turns out that consistent propagation on constant electromagnetic backgrounds (in the absence of other fields) are possible only with $g=2$. However, the analysis in \cite{Cortese:2013lda} left open the possibility of other values for $g$ in more complicated theories with additional fields. Similar analysis has also been performed by imposing causality and no superluminal propagation in constant electromagnetic background leading to the same value $g=2$ \cite{Porrati:2009bs,Porrati:2010hm,Kulaxizi:2012xp}. Again, however more complicated cases with additional fields have not been yet analysed from this perspective.

In fact, there are some exceptions to this natural value of the gyromagnetic factor. In \cite{HIOY:84}, it was shown that the massive spin-two particles, arising from compactification of the five-dimensional Einstein-Hilbert action, have $g=1$. The same value was obtained via soft-theorem compactifications of $d+1$ dimensional theory to $d$ dimension in \cite{Marotta:2019cip}. The D0 branes in Type IIA theory, which are the KK modes associated to the circle compactification of $D=11$ SUGRA, were also explicitly shown to have $g=1$ in \cite{9801072}.

In this draft, motivated by these results, we closely analyse the gyromagnetic factor for massive string excitations with respect to the $U(1)$ gauge fields arising from both the Graviton and Kalb Ramond fields upon compactification. We clarify how the value $g=2$ for open string states in 10 dimension is a simple consequence of conservation of the corresponding vertex operator. For modes with vanishing winding charge, the $g=1$ value for closed string states, turns out to arise as a simple consequence of the structure of the Graviton vertex operator that is written as a double-copy of open string vertex operators. Thus, for all string excitations with a field theoretic interpretation, the value $g=1$ simply follows from the consistency of minimal coupling before the reduction.

More specifically, we shall consider the compactification of Bosonic, type II and Heterotic string theories on general toroidal backgrounds in the presence of the Graviton and Kalb Ramond fields. We shall keep the internal components of these fields (which are related to moduli) to non-zero but arbitrary constant values. Following the method used in \cite{9405117} to determine the Hamiltonian of the bosonic string compactified on circles, we shall read the gyromagnetic ratios from the expressions of interacting Hamiltonians which describe the interaction between gauge fields and the massive HS fields under compactification. The same results will also be obtained from an amplitude calculation performed by considering a momentum  expansion of the massless string vertex. The final result about the electromagnetic coupling of string excitations to the first order in the field expansion and derivatives, can be written in terms of the minimal coupling plus non-minimal terms proportional to the internal spin operators $S_{R;L}^{\mu\nu}$, and it is invariant in form for all the string theories above considered. It turns out to be:
\begin{multline}\label{VAB0}    \mathcal{V}_{(A,B)\Phi\Phi}\simeq\Big\langle\Phi\Big|(p_L^a+p_R^a)A_a\cdot p-\frac{1}{2} F^A_{\mu\nu;
 a} (p_{L}^aS_R^{\mu\nu} +p_{R}^a S_L^{\mu\nu} )\\+(p_L^a-p_R^a)B_a\cdot p-\frac{1}{2}F^B_{\mu\nu;a} (p_{L}^a S_R^{\mu\nu} -p_{R}^aS_L^{\mu\nu} )\Big|\Phi\Big\rangle
\end{multline}
where $F_{\mu\nu;\,a}^{A;B}$ are the field strengths of the gauge fields arising from the metric and Kalb-Ramond compactification respectively, while $p_{R;L}^a$ are the compact momenta and $p$ is the momentum of the higher-spin states.

The states with field theoretic interpretation turn out to have $g=1$ as mentioned above. However, the compactification also gives rise to states characterized by Kaluza Klein and Winding charges. These are the charges with respect to the gauge fields arising due to the compactification of the Graviton and Kalb Ramond fields respectively. It turns out that the vanishing of either of these charges corresponds to states with $g=1$. On the other hand, the general twisted sector states have gyromagnetic ratios which depends upon the KK and Winding charges as well as spins. While these cannot be predicted by analysing minimal gravitational coupling and its dimensional reduction, it is clear that their pattern is also fixed by the double-copy structure of the graviton vertex operator.

One subtlety associated with the higher spin fields is regarding their symmetry properties. The general string states have mixed symmetry polarisation tensors which are characterised by a generic Young Tableau diagram. However, in the literature, these states have not received much attention. Our analysis shall also be applicable to these states. In particular, we give explicit results for the mixed symmetry states described by Young tableau diagrams having two rows from first Regge trajectory of string spectrum. It turns out that there is one gyromagnetic coupling (and hence a gyromagnetic ratio) associated with each row of the Young Tableau diagram describing the corresponding string state.\footnote{For an interesting instance of two gyromagnetic ratios in the context of black holes in Heterotic string theory on torus, see \cite{Sen:1994eb}.} One of our main result is the following gyromagnetic factors
\begin{align}
    g_1^{(a)}&=\frac{2}{p_L^a-p_R^a}\frac{(\ell_R-k) p_L^a-(\ell_L-k) p_R^a}{\ell_R+\ell_L-2k}\,,\\
    g_2^{(a)}&=\frac{2}{p_L^a-p_R^a}\frac{-(\ell_R-k) p_R^a+(\ell_L-k) p_L^a}{\ell_R+\ell_L-2k}\,,
\end{align}
for the mixed-symmetry states in the first Regge trajectory of closed string theories with respect to the gauge boson obtained from compactification of the metric. In the above expression, the $\ell_L$ and $\ell_R$ denote the left and right spins respectively. The state is represented by an Young diagram with $\ell_L+\ell_R-k$ boxes in the first row and $k$ boxes in the second row (see figure \ref{Younghook}). The first gyromagnetic factor coupling is associated to the first row while the second to the second row.

The rest of the draft is organised as follows. In section \ref{sec:2}, we discuss some generalities about the gyromagnetic coupling and introduce some notations which will be useful in our computations. In section \ref{sec2:hamiltonian}, we compute the part of interacting Hamiltonian of the compactified Bosonic, type II and Heterotic theories which encode the information about the gyromagnetic ratios. In section \ref{sec:stringAmpl}, we shall directly compute the 3 point vertices of gauge fields and the massive string states under the compactification using the string amplitudes. In section \ref{Examples}, we shall read the $g$ factor for some massive states using the results of section \ref{sec2:hamiltonian} and \ref{sec:stringAmpl}. Finally, we shall conclude with some discussion in section \ref{s4}. Details of some computations will be given in appendices.

\section{Some generalities about gyromagnetic factor(s)}
\label{sec:2}
In this section, we shall review the concept of gyromagnetic ratio. It will be convenient for the purposes of the present draft to define the gyromagnetic ratio as a coupling constant appearing in the effective action. In particular, considering the electromagnetic coupling of a field $\Phi$, at the lowest derivative order we can write down two type of couplings: minimal coupling which defines the charge of the field $\Phi$ and it is therefore uniquely characterised by this property and non-minimal couplings proportional to the Electromagnetic tensor $F_{\mu\nu}$
\begin{align}\label{APhiPhi}
    \mathcal{V}_{A\Phi\Phi}\sim iqA_\mu\left[\Phi^*\cdot(\partial^{\mu}\Phi)-(\partial^\mu\Phi^*)\cdot\Phi\right]+\frac{i\alpha}{2} F_{\mu\nu}(\Phi^{\mu}\cdot\Phi^{\nu})\,.
\end{align}
Above we have been schematic with the ``$\,\cdot\,$'' implying certain index contractions among the field which in principle can have an arbitrary tableaux shape. We have also canonically normalised the kinetic term as
\begin{align}
    \mathcal{L}=\Phi^\star\cdot(\Box+m^2)\Phi\,,
\end{align}
assuming the index contraction is of weight $1$ in the permutation of the various indices. With these conventions one can define the gyromagnetic factor as
\begin{align}
    g\sim\left|\frac{\alpha}{q}\right|\,.
\end{align}
Note that the gyromagnetic ratio is ill-defined if the charge $q$ vanishes. However one can still define the coupling $\alpha$.
The above definitions can then be related to the magnetic moment of a particle in 4d. The main simplification is that in 4d all particles can be classified as totally symmetric fields which implies the existence of a unique non-minimal electromagnetic coupling of the type \eqref{APhiPhi}
\begin{align}
    \mathcal{V}_{A\Phi\Phi}^{N.M.}=\frac{i\alpha}2 F_{\mu\nu}\,\Phi^{*\,\mu\mu(s-1)}{\Phi^{\nu}}_{\mu(s-1)}\,,
\end{align}
Using a generating function notation $\Phi(u)=\tfrac{1}{s!}\Phi_{\mu(s)}u^{\mu(s)}$, where we have introduced the totally symmetric product of $u$'s as $u^{\mu(s)}=u^{\mu_1}\cdots u^{\mu_s}$, the above equation can also be equivalently written as
\begin{align}\label{gyroSymm}
    \mathcal{V}_{A\Phi\Phi}^{N.M.}=\frac{i\alpha}{4}\, F_{\mu\nu}\left\langle\Phi|S^{\mu\nu}|\Phi\right\rangle\,,
\end{align}
where we have conveniently introduced the spin operator $S^{\mu\nu}_u=u^\mu\partial_{u}^\nu-u^\nu\partial_{u}^\mu$ together with the inner product defined as:
\begin{align}
    \left\langle\Phi_1|\Phi_2\right\rangle=\exp\left(\partial_{u_1}\cdot\partial_{u_2}\right)\ \Phi_1(u_1)\Phi_2(u_2)\Big|_{u_i=0}\,.\label{2.6}
\end{align}
The manifest appearence of the spin operator clarifies the relation between the angular momentum and the associated magnetic moment. E.g., for a classical massive electrically charged particle with mass $m$ and charge $q$, the magnitude of the magnetic moment \textbf{$\bm\mu$} and the angular momentum $\bm{L}$ are related as (see, e.g. \cite{Jackson:1998nia})
\be
\textbf{$\bm\mu$} =\frac{\alpha}{2m}\,\bf{L}\,.
\ee
Considering the generic dimensional case the story for totally symmetric representation is unchanged. However, in $d>4$ there exist more general representations of the Lorentz group and the number of non-minimal electromagnetic couplings can increase. In particular we have as many gyromagnetic factors as the number of rows of the fields. This also correspond to the number of spin operators which is equal to the number of rows. For concreteness, representing a generic mixed-symmetry field as a generating function in terms of auxiliary vector variables $u_i$
\begin{align}
    \Phi(u_i)=\frac1{s_1!\cdots s_n!}\phi_{\mu_1(s_1)\mu_2(s_2)\ldots\mu_n(s_n)}u_1^{\mu_1(s_1)}\cdots u_n^{\mu_1(s_n)}\,,
\end{align}
subject to the irreducibility conditions $u_{i}\cdot\partial_{u_{i+k}}\Phi=0$ for all $i$ and $k>0$, we can define the spin-operators:\footnote{The label $u_i$ in $S_{u_i}$ will be replaced by a simple index $i$ when no ambiguity can arise.}
\begin{align}
    S_{u_i}^{\mu\nu}=u_i^{\mu}\partial_{u_i}^{\nu}-u_i^{\nu}\partial_{u_i}^{\mu}\,.
\end{align}
One is then naturally led to the following basis of gyromagnetic factors:
\begin{align}\label{generalGyro}
    \mathcal{V}^{(j)}_{A\Phi\Phi}=\frac{i}4\alpha^{(j)}\,F_{\mu\nu}\left\langle\Phi|S_{j}^{\mu\nu}|\Phi\right\rangle\,.
\end{align}
A natural generalisation of the totally-symmetric gyromagnetic factor to mixed symmetry fields is obviously the one that involves the total spin operator
\begin{align}\label{standardGyro}
    \mathcal{V}_{A\Phi\Phi}=\frac{i}4\alpha\,F_{\mu\nu}\Big\langle\Phi\Big|\underbrace{\sum_{j=1}^nS_{j}^{\mu\nu}}_{S^{\mu\nu}}\Big|\Phi\Big\rangle\,.
\end{align}
In the following we shall focus on extracting the above coefficients $\alpha^{(j)}$ in string compactifications.

\subsection{Symbols in String Theory}

In this work, we shall often need to translate results from String Theory to extract the gyromagnetic factor. In order to do so, it is convenient to introduce a dictionary between string calculations in terms of oscillators and the auxiliary variables $u_i$ introduced above.

While in bosonic string theory one works with states of the form
\begin{align}
|\phi\rangle&= {\cal N}_0\ \phi_{\mu_1(s_1) \dots \mu_p(s_p)\bar{\mu}_1(\bar{s}_1)\dots  \bar{\mu}_q (\bar{s}_q)}\, \alpha^{\mu_1(s_1)}_{-n_1} \dots\alpha^{\mu_p(s_p)}_{-n_{p}}\,\bar{\alpha}^{\bar{\mu}_1(\bar{s}_1)}_{-\bar{n}_1}\dots\bar{\alpha}^{\bar{\mu}_{q}(\bar{s}_q)}_{-\bar{n}_{q}}\, |0,\,\bar{0},\,p\rangle
\label{string_states}
\end{align}
expressed in term of $\alpha$-oscillators, it is often more convenient to work with auxiliary commuting variables $w_i$ sometime referred to as symbols of the oscillators. Performing this step is actually extremely simple and can be obtained by replacing the $\alpha$-oscillators $\alpha_{-n}$ with commuting variables $w_n$ as
\begin{align}\label{mapping}
   |\phi\rangle\to \frac1{s_1!\cdots s_p! \bar{s}_1!\cdots \bar{s}_p!}\,\phi_{\mu_1(s_1) \dots \mu_p(s_p)\bar{\mu}_1(\bar{s}_1)\dots  \bar{\mu}_q (\bar{s}_q)}\, w^{\mu_1(s_1)}_{n_1} \dots w^{\mu_p(s_p)}_{n_{p}}\,\bar{w}^{\bar{\mu}_1(\bar{s}_1)}_{\bar{n}_1}\dots\bar{w}^{\bar{\mu}_{q}(\bar{s}_q)}_{\bar{n}_{q}}
\end{align}
The operator product governing the $\alpha$ oscillators can then be implemented as a certain differential operators on the symbols. For instance, the inner product then takes the form \eqref{2.6} where $u$ and $\bar{u}$ should be considered as different variables. It is also important to stress that the auxiliary variables $w_n$ are dummy variables and can be changed as needed.

In the following it will be useful to also remove the polarisation tensor replacing it with a product of vector polarisations. Introducing the vector polarisations $u_n$ and $\bar{u}_n$ one then arrives to the following representation for the states in \eqref{string_states}
\begin{align}\label{basis}
    \frac1{s_1!\cdots s_p! \bar{s}_1!\cdots \bar{s}_p!}\, (u_{n_1}\cdot w_{n_1})^{s_1} \dots (u_{n_p}\cdot w_{n_p})^{s_p}\,(\bar{u}_{\bar{n}_1}\cdot\bar{w}_{\bar{n}_1})^{\bar{s}_1}\dots(\bar{u}_{\bar{n}_q}\cdot\bar{w}_{\bar{n}_q})^{\bar{s}_q}\,.
\end{align}
Focusing on the first Regge trajectory we have only $\alpha_{-1}$ oscillators and the general closed string operator reads:
\begin{align}
\phi_{\ell_R+\ell_L}={\cal N}_{\ell_R,\,\ell_L}\,(u\cdot w)^{\ell_R}\,(u\cdot \bar{w})^{\ell_L}\,.
\end{align}
Here, however, we have a reducible representation. It turns out that using the same auxiliary variables, we can also implement conveniently Young projections by working with appropriate polynomials which implement the Young projection conditions.
For instance, mixed symmetry states associated to the Young-Tableaux shown in Figure \ref{Younghook}, are constructed in appendix \ref{Young}, by starting from the states of the form in Eq. \eqref{basis}. Here, we only write the result
\begin{align}
    \phi_{\ell_L+\ell_R-k,k}={\mathcal{N}}_{\ell_L+\ell_R-k,k}(u\cdot w)^{\ell_1-k} ({u}\cdot \bar{w} )^{\ell_2-k}\left(u\cdot w \, \bar{u}\cdot \bar{w}- {u}\cdot \bar{w}\,  \bar{u}\cdot w\right)^k\,,
\end{align}
with $k$ an integer number $k\leq {\rm Min}\{\ell_1,\,\ell_2\}$.
The expression of such states in terms of oscillators is easily obtained from the mapping defined in equation \eqref{mapping}.

In this work it will prove very convenient to work with the above polynomials in contractions of the auxiliary variables in order to extract the gyromagnetic ratio from the closed string amplitude. We would also like to stress that similar mappings can also be defined for Heterotic or Type II theories.

\section{Hamiltonian approach}
\label{sec2:hamiltonian}
String theories are naturally defined in the critical space-time dimensions where the conformal anomaly is vanishing. Their spectrum contains massless states identified  with  the gauge bosons of the fundamental  interactions and  an infinite tower of massive higher spin fields with mass proportional to the inverse of the string slope $\alpha'$. They provide  a quantum description of the interactions of such massive high-spin particles with massless and massive states and, therefore, are  consistent higher spin theories with an infinite number of fields.

Extracting interesting information about the string theory requires the compactification of several spatial dimensions. In the compactification, the initial  Lorenz group $SO(d-1,1)$ is broken  and new massless gauge-fields appear in the spectrum. These are the $d$-dimensional massless tensor fields having one index transforming as a vector under the unbroken $SO(d-D-1,1)$ and the other ones pointing along the compact directions. In the following, we shall focus  on the $U(1)$ gauge fields resulting from the compactification of the  metric and Kald-Ramond field and we shall compute the cubic interactions of  such gauge fields with  the higher spin fields compactified on a generic  torus $T^D$. From these cubic couplings, we can read the gyromagnetic ratios of the high-spin states charged with respect to these $U(1)$-fields.

In this section, we shall follow the Hamiltonian approach which gives a convenient way to extract the gyromagnetic coupling of massive string states \cite{9405117}. In this method, we shall compute the expectation value of the Hamiltonian describing the interaction between the gauge fields and the world-sheet fields. This Hamiltonian, in the case of Bosonic string theory compactified on circles, was considered in \cite{9405117}.  We  extend such a  result to the Bosonic, Type II and Heterotic string theories compactified on a generic $D$-dimensional compact torus parametrized  by the constant background components of the metric and the antisymmetric tensor. Before moving further, we note that for comparing with the standard expression of Hamiltonian containing the gyromagnetic coupling, it is convenient to multiply the expression of the Hamiltonian with $\ell/(2\pi \alpha' m)$ and work with the following rescaled Hamiltonian\footnote{One way to see this is to note that the zero mode of the Hamiltonian in equation \eqref{pim2} contains the term $ H_0=\frac{\alpha'\pi }{\ell}\,\hat{p}^2+\dots $ with $\hat{p}^\mu$ being the momentum operator along the non compact directions. Evaluating its expectation value for an external state having momenta $(p_0,\vec{p})$ with mass $m$ and multiplying by $\ell/(2\pi \alpha' m)$ gives the standard expression $\vec{p}^2/2m$ of the energy of the free charged particle of mass $m$. This is also in the agreement with the expressions in the non relativistic limit.} \cite{9405117} 
\begin{eqnarray}
 {\cal H}_I\equiv \frac{\ell}{2\pi \alpha' m}\, H_I. \label{32}
\end{eqnarray}

\subsection{Details of Compact Toroidal Background}
\label{compbasis}
The manifold on which we shall analyse the string theories are generic toroidal manifolds $T^D$. We shall denote with $\mu, \nu=0,\dots d-D-1$ the non compact directions, and with $i,j=1,\dots D$ the compact ones. The $d$ will be 26 for the Bosonic theory and 10 for the superstring theories. Since torus is a flat manifold, we can parametrize it so that the metric $g_{ij}$ on it is constant and the compact coordinates $X^i$ have the period $2\pi\sqrt{\alpha'}$, i.e.
\be
X^i(\uptau,\sigma +\ell) = X^i(\uptau,\sigma) + 2\pi\sqrt{\alpha'}\ n^i\qquad,\quad n^i\in \mathbb Z
\ee
Here, $\ell$ can be either $\pi$ or $2\pi$ depending on the adopted conventions. In the above equation, the more general boundary conditions are allowed for the compact directions because the points $X^i$ and $X^i+2\pi\sqrt{\alpha'} n^i$ describe the same point in space-time. In contrast, for the non compact directions, the boundary conditions are 
\begin{eqnarray}
X^\mu(\uptau, \sigma+\ell)&=& X^\mu (\uptau,\sigma )
\end{eqnarray}
Instead of $g_{ij}$, it is convenient to work with the standard Euclidean metric $\delta_{ab}$. This can be done by introducing a constant vielbein $e_i^a$ on the torus via  
\begin{eqnarray} 
g_{ij}= e_i^{a}\, \, e_j^b\delta_{ab}\label{7n}
\end{eqnarray} 
With this, the boundary condition for the compact directions can be written as
\begin{eqnarray}
X^a(\uptau,\sigma+\ell)&=&  X^a(\uptau,\sigma) +2\pi\sqrt{\alpha'} L^a\quad;\qquad L^a=\sum_{i=1}^Dn^ie_i^a~~;~~n^i\in \mathbb Z\label{bghfn}
\end{eqnarray}
where,
\be
X^a(\sigma,\uptau)=e_i^a X^i(\sigma,\uptau).
\ee
The $\textbf{e}_i\equiv\{e_i^a\}$ can be interpreted to be $D$ linearly independent vectors which generate a $D$ dimensional lattice $\Lambda_D$. The $\textbf{L}\equiv\{L^a\}$ in \eqref{bghfn} are lattice vectors, i.e., $\textbf{L}\in \Lambda_D$. The compactified torus can now be viewed as the quotient of  $\mathbb{R}^D$ by $2\pi \Lambda_D$\cite{9401139}, i.e.:
\be
T^D=\f{\mathbb{R}^D}{2\pi \Lambda_D}
\ee
We can also define a dual lattice by introducing vectors $\textbf{e}^{*i}=\{e^{*i}_a\}$ via\footnote{The identity \eqref{7n} also gives $e^{*j}_cg_{ij}e^{*i}_d=\sum_{a=1}^De_i^ae^{*i}_d \,\delta_{ab}\, e_j^b e^{*j}_c=\delta_{dc}$, showing that the D-vectors ${\bf e}_a^*\equiv(e^{*i}_a)$ form an orthonormal basis.} :
\begin{eqnarray}
e^{*i}_b\,e^a_i=\delta^a_b~~;~~ e_i^a\,e^{*j}_a=\delta^j_i~~;~~\sum_{a=1}^D e^{*j}_a\,e^{*i}_a=g^{ij}
\end{eqnarray}
We shall be mostly concerned with the zero modes of the compactified fields which correspond to the massless fields in the lower dimension. For the metric, we express the general compactification  ansatz as
\begin{eqnarray}
G_{MN} =\Bigg(\begin{array}{ccc}
g_{\mu\nu}+g_{kl} A_\mu^k\,A_\nu^l&& A_{\mu j}\,\\ 
A_{i\nu}\,&& g_{ij}\end{array}\Bigg) ~~;~~G^{MN}=\Bigg(\begin{array}{ccc}
g^{\mu\nu }&& -A^{\mu j}\\ 
-A^{i \nu} && g^{ij} +g^{\rho\sigma}A_\rho^i\,A^{ j}_\sigma\end{array}\Bigg)\label{3.5.4}
\end{eqnarray}
where we are assuming that all the quantities depend only on the non compact directions $X^\mu$ of the space-time (since we are only interested in the zero-modes).
Here $A_\mu^i$ are vector fields
in the lower dimension. The $g_{\mu\nu}$ is the non compact space-time metric and $g_{ij}$ are the scalars in the lower dimensional  theory. Their expectation values can be related to the moduli of the compatified torus $T^D$.The background values of $g_{\mu\nu}$ and $g_{ij}$ will be used to lower and raise the non compact and internal indices respectively.

\vspace*{.07in}Similarly, for the Kalb-Ramond field, we have
\begin{eqnarray}
B_{MN}=\Bigg(\begin{array}{cc}
B_{\mu\nu}&B_{\mu j}\\
{B_{i\nu }} &B_{ij}\end{array}\Bigg)\qquad,\quad B_{\mu i}=-B_{i\mu}\label{bmn}
\end{eqnarray}
The $B_{\mu i}$ denote set of another vector fields in the lower dimension. $B_{\mu\nu}$ is an antisymmetric 2nd rank tensor field in the lower dimension. It will be taken to be vanishing in most of our discussion. The $B_{ij}$ denote other set of scalar fields in the lower dimension. Again, the expectation values of $B_{ij}$ can be related to the moduli of the compactified manifold $T^D$.

\vspace*{.07in}The infinitesimal coordinate transformation along the compact directions $X^i\rightarrow X^i+\xi^i$, where $\xi$ depends only on the non-compact coordinates, changes the vector field $A_{\mu i}$ as
\begin{eqnarray}
\delta A_{\mu i}(X^\mu)= g_{ij} \partial_\mu\xi^i\ =\partial_\mu\xi_j \qquad\Leftrightarrow \quad\delta A_{\mu a}=\partial_\mu\xi_a\quad;\qquad A_{\mu a} =e^{*i}_a A_{\mu i}
\end{eqnarray}
This is a $U(1)$ gauge transformation acting on the vector field $A_{\mu a}$. For our purposes, it will be enough to take the field strength of this gauge field to be a constant\footnote{If the field strength is not constant, there will be $\alpha'$ dependent corrections in \eqref{fieldstrenght} \cite{9405117}. However, these corrections do not contribute to the gyromagnetic ratio since the terms with higher orders in $\alpha'$ correspond to the higher derivative terms. },
in which case, we can write
\be
A_{\mu a}=-\frac{1}{2}F_{\mu\nu; a}\,X^\nu\label{fieldstrenght}.
\ee
Similarly, the gauge transformation for the Kalb Ramond field is given by
\begin{eqnarray}
\delta B_{MN}=\partial_M\Lambda_N-\partial_N\Lambda_M\label{3}
\end{eqnarray}
Specializing this to a gauge parameter $\Lambda_i$ depending only on the non compact coordinates gives
\begin{eqnarray}
\delta B_{\mu i}=\partial_\mu \Lambda_i
\end{eqnarray}	  
This is also a $U(1)$ gauge transformation for the gauge field $B_{\mu i}$. As above, we shall take the field strength of this gauge field also to be a constant.

As mentioned above, the metric $g_{ij}$ will be taken to have the constant expectation value given by \eqref{7n}. For the other fields, we shall assume the following background values
\begin{eqnarray} 
g_{\mu\nu}=\eta_{\mu\nu}~~~~~~;~~B_{\mu \nu} =0\quad;\quad B_{ij}=\mbox{const}\label{7}
\end{eqnarray} 
with $\eta_{\mu\nu}$ the Minkowski (mostly plus) metric for the non compact directions.

\subsection{Closed Bosonic strings on $T^{D}$}
\label{bosonic}

The sigma model action in 26 dimensions in the presence of the space-time metric $G_{MN}(X)$ and an antisymmetric tensor $B_{MN}(X)$ is given by \cite{GSWI}
\begin{eqnarray}
S&=&-\frac{1}{4\pi \alpha'} \int d^2\sigma \big[G_{MN}(X)\, \eta^{\alpha\beta} \partial_\alpha X^M\,\partial_\beta X^N+B_{MN}(X) \epsilon^{\alpha\beta} \partial_\alpha X^M\,\partial_\beta X^N\Big]\label{4.1.1in}
\end{eqnarray} 
where, the space-time indices $M,N$ run over $0,\dots d-1$ with $d=26$. The world-sheet metric is given by $\eta_{\alpha\beta}=(-1,1)$ and our convention for the Levi civita tensor is $\epsilon^{01}=-\epsilon_{01}=1$. Also, $\sigma^\alpha\equiv (\tau,\,\sigma)$. 

\vspace*{.07in} The canonically conjugate momenta of $X^M$ is given by
\begin{eqnarray}
\Pi_M= \frac{\delta L}{\delta\partial_0 X^M}=\frac{1}{2\pi\alpha'}\Big[G_{MN}\,\partial_0X^N-B_{MN} \,\partial_\sigma X^N\Big]\label{pim}
\end{eqnarray} 
The Hamiltonian corresponding to \eqref{4.1.1in} is computed to be
\be
H&=&\int_0^\ell d\sigma\Bigl[\Pi_M\p_0X^M-\mathcal{L}\Bigl]\non\\
&=&\f{1}{2}\int_0^\ell d\sigma\biggl[(2\pi\alpha')\Pi_M G^{MN}\Pi_N + 2B_{MN}G^{MP}\Pi_P\partial_\sigma X^N\non\\
&&\hspace*{.4in}+\ \ \f{1}{2\pi\alpha'}\p_\sigma X^M\Bigl(G_{MN}+G^{PQ}B_{PM}B_{QN}\Bigl)\partial_\sigma X^N\biggl]
\label{pim2}
\ee
For analysing the gyromagnetic couplings of the massive string states with respect to the gauge fields $A^i_\mu$ and $B^i_{\mu}$, we shall need the terms linear in the gauge fields in the Hamiltonian describing the interaction between these gauge fields and the world-sheet fields $X^i$ and $X^\mu$. This is easily obtained, for the background defined by \eqref{7}, by substituting the compactification ansatz \eqref{3.5.4} and \eqref{bmn} in the expression \eqref{pim2}
\begin{eqnarray}
H_I
&=&-\f{1}{2\pi\alpha'}\int_0^{\ell}d\sigma\biggl[A^{\mu i}\biggl((2\pi\alpha')^2\Pi_\mu\Pi_i+(2\pi\alpha')B_{ij}\Pi_\mu\p_\sigma X^j-g_{\mu \nu}g_{ij}\p_\sigma X^\nu\p_\sigma X^j\biggl)\non\\
&&\hspace*{.3in}+ \ B_{\mu i}\Bigl(-(2\pi\alpha')\Pi^\mu\p_\sigma X^i+(2\pi\alpha')\Pi^i \p_\sigma X^\mu+B_{jk}g^{ ki}\p_\sigma X^\mu \p_\sigma X^j\Bigl)\biggl]\label{4.1.13}
\end{eqnarray} 
where, 
\be
\Pi_\mu
&\equiv& P_\mu+\f{1}{2\pi\alpha'}\Bigl[A_{i\mu}\p_0X^i-B_{\mu i}\p_\sigma X^i-B_{\mu\nu}\p_\sigma X^\nu\Bigl]+O(A^2, B^2, AB)\label{amud}\\[.3cm]
\Pi_i
&\equiv&P_i+\f{1}{2\pi\alpha'}\Bigl[A_{i\mu} \p_0X^\mu+B_{\mu i}\p_\sigma X^\mu\Bigl]+O(A^2, B^2, AB)\label{amud1}
\ee
with
\be
P_\mu= \f{1}{2\pi\alpha'}g_{\mu\nu}\p_0X^\nu\qquad,\quad P_i = \f{1}{2\pi\alpha'}\Bigl[g_{ij}\p_0X^j-B_{ ij}\p_\sigma X^j\Bigl].\label{Pmupm}
\ee
In the Hamiltonian \eqref{4.1.13}, being interested in the terms which are linear in the gauge fields, we can replace $\Pi_{\mu,i}$ with $P_{\mu,i}$. Furthermore by explicitly evaluating these quantities on the solution of the equations of motion given in equation \eqref{13}, we find that all the dependence on the moduli $B_{ij}$ disappears  and we can express \eqref{4.1.13} as 
\be
H_I&=&H_I^A+H_I^B
\ee
where,
\be
H_I^A&=& \frac{1}{2\pi\alpha'}F^A_{\mu\rho;\,i}\int_0^{\ell} d\sigma {\cal{X}}^\rho\Big(\p_+{\cal{X}}^\mu\p_-X^i+\p_-{\mathcal{X}}^\mu\p_+X^i\Big)\non\\[.2cm]
H_I^B&=& \frac{1}{2\pi\alpha'}F^B_{\mu\rho;\,i} \int_0^{\ell} d\sigma\ \mathcal{X}^{\rho} \Big(\p_+\mathcal X^\mu\p_-X^i-\p_-\mathcal{X}^\mu \p_+X^i\Bigl) \label{3.5.14dfg}
\end{eqnarray}
The expectation value for $H^A_I$ which describes the interaction of the gauge field $A^\mu_i$ with the world-sheet fields is computed to be
\begin{eqnarray}
\langle\phi|{\cal H}_I^A|\phi\rangle
&=&-\f{1}{2 m} F^A_{\mu\nu;\,i} \Bigl\langle\phi\Bigl|\f{1}{2}L^{\mu\nu}\underbrace{\bigl( p_R^i+p^i_L\bigl)}_{Q^i}\ +\ p^i_RS_L^{\mu\nu}  \:+\  p^i_LS_R^{\mu\nu}\Bigl|\phi\Bigl\rangle\label{hia}
\end{eqnarray}
The $p_L$ and $p_R$ in the above expression denote the left and right momenta associated with string compactification and are defined in equation \eqref{16}. The $L^{\mu\nu}$ denotes the orbital angular momentum and $S^{\mu\nu}=S_L^{\mu\nu}+S_R^{\mu\nu}$ denotes the spin angular momentum with
\begin{eqnarray}
S_R^{\mu\nu}=-i\sum_{n=1}^\infty \frac{1}{n} \bigl(\alpha^\mu_{-n} \alpha^\nu_n-\alpha^\nu_{-n} \alpha^\mu_n\bigl)\quad,\qquad
S_L^{\mu\nu}=-i\sum_{n=1}^\infty \frac{1}{n} \bigl(\tilde{\alpha}^\mu_{-n} \tilde{\alpha}^\nu_n-\tilde{\alpha}^\nu_{-n} \tilde{\alpha}^\mu_n\bigl)\label{37}
\end{eqnarray}
In the following we shall identify these two operators with the names ``right'' and ``left'' spin operators. In equation \eqref{hia}, the {\em Kaluza-Klein} charges of the massive higher spin states are given by \footnote{{In Ref.\cite{Marotta:2019cip} the KK-charge is taken equal to the compact momentum $p_z=n/R$. In Eq. \eqref{38}, when $p_R=p_L$, the charge is $(p_R+p_L)_ie^{*i}_a=\frac{n_i}{\sqrt{\alpha'}} e^{*i}_a$. The choice $e^{*i}_a=\delta^i_a\,\frac{\sqrt{\alpha'}}{R_a}$ and $e_i^a=\delta_i^a\,\frac{R_a}{\sqrt{\alpha'}}$ matches the two charges and gives $X^a\equiv X^a+2\pi R_a n^a$ which is the identification of the compact coordinates used in the above reference.} }\cite{9405117} 
\begin{eqnarray}
Q^a=\frac{1}{2\pi \alpha'}\int_0^\ell d\sigma \partial_\tau X^a=e^{*i}_b(p_R+p_L)_i\delta^{ab}\label{38}
\end{eqnarray}
In a similar manner, for the gauge field $B_{\mu i}$, we find
\begin{eqnarray}
\langle\phi|{\cal H}_I^B|\phi\rangle
&=&-\f{1}{2m} F^B_{\mu\nu;\,i} \langle\phi |\f{1}{2}L^{\mu\nu} \underbrace{\bigl( p_R^i-p^i_L\bigl)}_{{\cal Q}^i}\ -\ S_R^{\mu\nu} p^i_L \:+\ S_L^{\mu\nu} p^i_R |\phi \rangle\label{hib}
\end{eqnarray}
with the {\em winding} charge of the $U(1)$ field given by \cite{9405117}:
\begin{eqnarray}
{\cal Q}^a=\frac{1}{2\pi\alpha'}\int_0^\ell d\sigma \partial_\sigma X^i \,e_{i}^a= (p_L-p_R)_i\,e^{*i}_b\,\delta^{ab}\label{50}
\end{eqnarray}
We notice that higher spin states with $p_R=p_L$ or $p_R=-p_L$ have  null charges ${\cal Q}_a$ or  $Q_a$ respectively. From equations \eqref{hia} and \eqref{hib}, we see that the gyromagnetic ratios of these particles are $g=1$ with respect to one gauge field. Note also that when this happens the same massive field is uncharged with respect to the other gauge field being only coupled non-minimally to it with a coupling proportional to the difference of the left and right spin operators. Further comments about these special cases will be given in Section \ref{Examples}.

\vspace*{.07in} It is worth noting that the expression \eqref{hib} can be obtained from equation \eqref{hia} by exchanging the two gauge fields and changing the sign of the compact momentum $p_{L}^i$. 
The two charges, instead, are transformed into each other as $(p_L,\,p_R)\leftrightarrow  (p_L,\,-p_R)$. In the case $B_{ij}=0$, this transformation is the $D$-dimensional analogous of the $R\rightarrow 1/R$-duality of the string theory compactified on a circle of radius $R$ \cite{9401139}.

\subsection{Type II and Heterotic strings on $ T^{D}$}
In this subsection, we consider the closed superstring theories and generalize the expression of the interacting Hamiltonian of the bosonic string theory obtained in the previous subsection to bosonic and fermionic states of the superstring theories. We start by considering the type II string theory compactified on the torus $T^D$ with $D=9-d$. After this, we shall extend this analysis to the Heterotic theory. Our approach will be exactly the same as in the case of the bosonic string theory. Thus, the starting point is the action of the $(1,1)$ supersymmetric sigma model in 10 dimensions in a generic non constant background \cite{callan1, Maharana, Hull&Witten:1985} (see appendix \ref{apps1} for details)
\begin{eqnarray}
S&=&\frac{1}{4\pi\alpha'} \int d^2\sigma \bigg[4G_{MN}\partial_+X^M\partial_-X^N+4B_{MN}\partial_+X^M\partial_-X^N +2iG_{MN}\psi_+^M\tilde{\nabla}_-\psi_+^N \nonumber\\
&&\hspace*{.87in}+\ 2iG_{MN} \psi_-^N\tilde{\nabla}_+\psi_-^M+\frac{1}{2}\tilde{R}_{MNPQ}\psi_+^M\psi_+^N\psi_-^P\psi_-^Q\bigg]\label{4.3.45}
\end{eqnarray}
where, the covariant derivatives $\tilde\nabla_\pm$ are defined by
\be
\tilde\nabla_\pm\psi^M_\mp=\p_\pm\psi_\mp^M+\tilde\Gamma^M_{\pm\,PQ}\psi_\mp^P\ \p_\pm X^Q\quad,\qquad \tilde\Gamma^M_{\pm PQ}=\Gamma^M_{\;\;PQ}\pm\f{1}{2}H^M_{\;\;PQ}
\ee
The $\tilde{\Gamma}^P_{\pm MN}$ are the connections with a totally antisymmetric torsion. The $\tilde R_{MNPQ}$ are given by
\begin{eqnarray}
\tilde R_{MNPQ}=R_{MNPQ}+\f{1}{2} \nabla_PH_{MNQ}-\f{1}{2} \nabla_QH_{MNP}+\frac{1}{4} H_{MRP}H^R_{\;\;QN}-\frac{1}{4} H_{MRQ}H^R_{\;\;PN}
\end{eqnarray}
The conjugate momenta for $X^M$ are computed to be
\begin{eqnarray}
\Pi_M&=& \frac{1}{2\pi\alpha'}\Big[G_{MN}\partial_0X^N-B_{MN}\partial_\sigma X^N-S_M\Big]
\label{pim1ws}
\end{eqnarray} 
where,
\be
S_M = -\f{i}{2}G_{PQ}\biggl[\tilde\Gamma^P_{-NM}\psi_+^Q\psi_+^N+\tilde\Gamma^P_{+NM}\psi_-^Q\psi_-^N\biggl]\label{4.5.tt}
\ee
The conjugate momenta corresponding to $\psi_\pm^M$ are given by
\be
\tau_\pm^M&=& \frac{\delta L}{\delta\partial_0 \psi_\pm^M}
\ =\ \frac{i}{4\pi\alpha'}G_{MN}\psi_\pm^N
\label{pim1wsq}
\ee
The Hamiltonian is computed to be
\be
H 
&=&\f{1}{4\pi\alpha'}\int_0^\ell d\sigma\Bigg[(2\pi\alpha')^2G^{MN}\Pi_M\Pi_N+2(2\pi\alpha')G^{MN}\Pi_MS_N+2(2\pi\alpha')G^{MN}\Pi_MB_{NP}\p_\sigma X^P\non\\
&&+2G^{MN}B_{MP}\p_\sigma X^PS_N+G^{MN}B_{MP}B_{NQ}\p_\sigma X^P\p_\sigma X^Q+G^{MN}S_MS_N+G_{MN}\p_\sigma X^M\p_\sigma X^N\non\\
&&-iG_{MN}\Bigl(   \psi^N_-\p_\sigma\psi^M_--\psi^N_+\p_\sigma\psi^M_+\Bigl)+2T_P\p_\sigma X^P-\frac{1}{2}\tilde R_{SMLP}\psi_+^S\psi_+^M\psi_-^L\psi_-^P\Bigg]\label{66}
\ee
where, we have defined 
\be
T_P\equiv \f{i}{2}G_{PQ}\Bigl(\tilde\Gamma^P_{-NM}\psi_+^Q\psi_+^N-\tilde\Gamma^P_{+NM}\psi_-^Q\psi_-^N\Bigl)\label{55tm}
\ee
We can now compute the part of Hamiltonian which describes the interaction between the string states and the gauge fields resulting from the compactification on $T^D$. For convenience, we organise them in two kind of terms: one in which strings interact with the external gauge field and the other in which they interact with their field strength. The terms describing the interaction of the world-sheet fields with only one  gauge field are given by
\be
H_1&=& \f{1}{2\pi\alpha'}\int_0^\ell d\sigma \biggl[A_{\mu i}\Bigl\{-(2\pi\alpha')^2\Pi^\mu\Pi^i-(2\pi\alpha')\Pi^\mu B^i_{~\;j}\p_\sigma X^j+\p_\sigma X^i \p_\sigma X^\mu\non\\
&&\hspace*{.5in}+\ \f{i}{2}\Bigl(  \psi^\mu_+\p_\sigma\psi^i_+ -\psi^\mu_-\p_\sigma\psi^i_-+ \psi^i_+\p_\sigma\psi^\mu_+ -\psi^i_-\p_\sigma\psi^\mu_-\Bigl) \Bigl\}\non\\
&&\hspace*{.5in}+\ B_{\mu i}\Bigl\{  (2\pi\alpha') \Pi^\mu \p_\sigma X^i+(2\pi\alpha')\Pi^i\p_\sigma X^\mu  +g^{ij}B_{jk}\p_\sigma X^k \p_\sigma  X^\mu\Bigl\}\biggl]\label{4.5.113ed}
\ee
and the terms describing the interaction with the field strength are given by
\be
H_2&=&- \f{i}{4}\int_0^\ell d\sigma \biggl[F^A_{\mu\nu;\,i}\Bigl\{\Pi^\mu\Psi_+^{\nu i}-\Pi^i\Psi_+^{\mu \nu}-\f{1}{2\pi\alpha'}\Bigl( B^i_{\;\;\;j}\p_\sigma X^j\Psi_+^{\mu\nu}+\p_\sigma X^\mu\Psi_-^{\nu i} -\p_\sigma X^i\Psi_-^{\mu \nu}\Bigl) \Bigl\}\non\\
&&+F^B_{\mu\nu; i}\Bigl\{ -2\Pi^\mu \Psi_-^{\nu i}-\Pi^i\Psi_-^{\mu\nu}  +\f{1}{2\pi\alpha'} \Bigl(-B^i_{\;\;j}\p_\sigma X^j\Psi_-^{\mu\nu}+2\p_\sigma X^\mu\Psi_+^{\nu i}+\p_\sigma X^i\Psi_+^{\mu \nu} \Bigl)\Bigl\}\biggl]
\label{59}
\ee
where, we defined $\Psi_\pm^{MN}=\psi_+^M\psi_+^N\pm \psi_-^M\psi_-^N$.

\vspace*{.07in}The expectation value of the interacting Hamiltonian between two generic string states is given by 
\begin{eqnarray}
\langle \phi|{\cal H}_I|\phi\rangle= \langle \phi|{\cal H}_1|\phi\rangle+\langle \phi|{\cal H}_2|\phi\rangle
\end{eqnarray}
We have divided equations \eqref{4.5.113ed} and \eqref{59} by the factor introduced in \eqref{32} to have the canonical normalization. The state $|\phi\rangle$ now also includes the fermionic oscillators along with the bosonic ones.

The calculation of the first term is exactly identical to the calculation of the bosonic case because the second line of \eqref{4.5.113ed} gives zero contribution when evaluated on the external states. This can be seen by inserting the mode expansion for the fields and noting that these terms change the level of the state and hence the inner product becomes zero. The remaining terms in \eqref{4.5.113ed} are exactly identical to the Bosonic Hamiltonian. 

\vspace*{.07in}Thus, we only need to focus on the second term of the above expression which involves $H_2$. As in the bosonic case, we can again replace the conjugate momenta $\Pi_m$ and $\Pi_\mu $ by $P_m$ and $P_\mu$ respectively, given in \eqref{Pmupm}, upto the linear order in the fields. Moreover, the terms proportional to $\psi_\pm^\mu\psi_\pm^m$ give vanishing contribution since they change the level of the states and the inner product vanishes due to orthogonality property. The expression can then be expressed in the form
\be
\langle \phi|{\cal H}_2|\phi\rangle
&=& \f{i}{4\pi\alpha'}F^A_{\mu\nu; i}\int_0^\ell d\sigma\bigl\langle \phi\bigl|\bigl\{\p_-X^i\psi_+^{\mu}\psi_+^\nu +\p_+ X^i\psi_-^{\mu}\psi_-^\nu \bigl\}\bigl|\phi\bigl\rangle\non\\[.2cm]
&&+\f{i}{4\pi\alpha'}F^B_{\mu\nu; i}\int_0^\ell d\sigma 
\bigl\langle\phi\bigl|\bigl\{ \p_-X^i\psi_+^{\mu}\psi_+^\nu  -\p_+ X^i\psi_-^\mu\psi_-^\nu \bigl\}\bigl|\phi\bigl\rangle\label{2,50}
\ee
We now consider the case $B_{\mu i}=0$. Using the mode expansions given earlier and performing the $\sigma$-integration, we find 
\be
\langle \phi|{\cal H}_2^A|\phi\rangle
&=& \frac{i}{2m} F^A_{\mu\nu; m}\sum_{r\in \mathbb Z+a}\Bigl\langle \phi\Bigl|\Bigl\{p^m_R\bar \psi^\mu_r\bar\psi^\nu_{-r}  +p^m_L \psi^\mu_r\psi^\nu_{-r} \Bigl\}\Bigl|\phi\Bigl\rangle\non\\
&=& -\f{1}{4m}F^{A\,a}_{\mu\nu}\Bigl\langle \phi\Bigl|\Bigl\{(p_R+p_L)_a(K_L^{\mu\nu}+K_R^{\mu\nu}) +(p_L-p_R)_a(K_R^{\mu\nu}-K_L^{\mu\nu}) \Bigl\}\Bigl|\phi\Bigl\rangle\non\\
\ee
where, 
\be
K_L^{\mu\nu}\, \equiv\, -\frac{i}{2}[\bar\psi_0^\mu,\,\bar\psi_0^\nu]\delta_{a,0}-i\sum_{r\in \mathbb N+a} (\bar\psi_{-r}^\mu\bar\psi_{r}^\nu-\bar\psi_{-r}^\nu\bar\psi_{r}^\mu)\Bigl)\nonumber\\
 K_R^{\mu\nu}\,\equiv\, -\frac{i}{2}[\psi_0^\mu,\,\psi_0^\nu]\delta_{a,0}-i\sum_{r\in \mathbb N+a}\Bigl(\psi_{-r}^\mu\psi_{r}^\nu-\psi_{-r}^\nu\psi_{r}^\mu  \Bigl)\label{75}
\ee
are the contribution to the angular momentum from the $\psi^M_\pm$ fields. This expression, when added to $\langle \mathcal{E}|{\cal H}_1^A|\mathcal{E}\rangle$ changes the bosonic result by the replacement of $S_{R,L}$ given in Eq.\eqref{37} and now denoted with $S_{R,L}^B$ with $S_{R,L}= S^B_{R,L}+K_{R,L}$, the spin operators in superstring theory, giving
\begin{eqnarray}
\langle\phi|{\cal H}_I^A|\phi\rangle
&=&-\f{1}{4 m}(F^A)^a_{\mu \nu}\,Q_a\bigl\langle \phi\bigl|(L^{\mu\nu}+S^{\mu\nu}) + \frac{ {\cal Q}_a}{Q_a} (S_R^{\mu\nu}-S_L^{\mu\nu}) \bigl| \phi\bigl\rangle\label{72}
\end{eqnarray}
where the charges are defined in equations \eqref{38} and \eqref{50}.

\vspace*{.07in}For the case of non zero $B_{\mu i}$, the calculation proceeds in the similar way. By looking at \eqref{2,50}, we find that the only difference in the calculation involving $B_{\mu i}$ as compared to $A_{\mu i}$ is in the sign in front the second term containing $\partial_+X^m$.
This corresponds to the replacement of $p_L$ with $-p_L$ and therefore in the exchange of  $Q_a$ with ${\cal Q}_a$ in the final expression, giving
\begin{eqnarray}
\langle \phi|{\cal H}_2^B|\phi\rangle= -\f{1}{4m}F^{A\,a}_{\mu\nu}\Bigl\langle \phi\Bigl|\Bigl\{(p_R-p_L)_a(K_L^{\mu\nu}+K_R^{\mu\nu}) +(p_L+p_R)_a(K_L^{\mu\nu}-K_R^{\mu\nu}) \Bigl\}\Bigl|\phi\Bigl\rangle
\end{eqnarray}
Again, when we add the contribution from ${\cal H}_1$, which is same as given in equation \eqref{hib}, it changes the bosonic result by the replacement $(S^B_{R,L})^{\mu\nu}\rightarrow  S^{\mu\nu}_{R,L}+K^{\mu\nu}_{R,L}$, the spin operators in supersymmetric theory.
Finally, we turn to the $SO(32)$ and $E_8\times E_8$ Heterotic theories compactified on the torus $T^D$. Again, we shall compute the Hamiltonian, giving the interaction between string world sheet fields and the gauge fields. The starting point will be the heterotic sigma model in ten dimensions in the presence of the background fields $G_{MN},\,B_{MN}$ and the gauge field $(A_M)^B_{\;\;C}$. The gauge group in the $E_8\times E_8$ or $SO(32)$ models could be equivalently  represented by fermionic or bosonic formulations. In the following, we follow the former approach and start from an action containing  $32$ Majorana-Weyl fermions $\lambda^A_-$ coupled to the background fields. The indices of $\lambda^A_-$ are lowered and raised by the metric $g_{AB}$. The sigma model action turns out to be \cite{callan1,Hull&Witten:1985,Sen1, Narain&Witten:1987} 
\be
S&=&\f{1}{4\pi\alpha'}\int d^2\sigma \Bigl[4G_{MN}\p_+X^M\p_-X^N+4B_{MN}\p_+X^M\p_-X^N+2iG_{MN}\psi_+^M\tilde\nabla_-\psi^N_+  \non\\[.2cm]
&&\hspace*{.5in} +2ig_{AB}\lambda_-^A\hat\nabla_+\lambda_-^B  +\f{1}{2}F_{MN;CD} \psi_+^M\psi_+^N\lambda^C_-\lambda_-^D\Bigl]\label{newacty1}
\ee
where the $\psi_+$ are left-moving fermions and $\tilde\nabla_-\psi^N_+$ is defined in the same way as in the type II case (see equation \eqref{gradpsiplus}) and 
\be
\hat\nabla_+\lambda_-^B=\p_+\lambda_-^B  +(\hat A_M)^B_{\;\;C}\lambda_-^C\p_+X^M\quad,\qquad (\hat A_M)^B_{\;\;C}=( A_M)^B_{\;\;C}+\f{1}{2}g^{BD}\p_Mg_{DC}
\ee
The field strength for the gauge field $(\hat A_M)^B_{\;\;C}$ is defined as 
\be
F_{MN;CD}=\p_M(\hat A_N)_{CD}-\p_N(\hat A_M)_{CD}+(\hat A_M)_{CB}(\hat A_N)^B_{\;\;D}-(\hat A_N)_{CB}(\hat A_M)^B_{\;\;D}
\ee
The conjugate momenta for $X^M$ are given by
\begin{eqnarray}
\Pi_M
=\frac{1}{2\pi\alpha'}\Big[G_{MN}\partial_0X^N-B_{MN}\partial_\sigma X^N-S_M\Big]
\label{pim1wsh}
\end{eqnarray} 
where, $S_M$ is now given by
\be
S_M = -\f{i}{2}G_{PQ}\tilde\Gamma^P_{-NM}\psi_+^Q\psi_+^N-\f{i}{2}g_{AB}(\hat A_M)^B_{\;\;C}\lambda_-^A\lambda_-^C\label{4.5.tth}
\ee
The conjugate momenta of $\psi_+^M$ is same as in the case of type II superstrings. For the $\lambda_-^A$, we have 
\be
\Pi_-^A&=& \frac{\delta L}{\delta\partial_0 \lambda_-^A}
=\frac{i}{4\pi\alpha'}g_{AB}\lambda_-^B
\label{pim1wsq1h}
\ee
The Hamiltonian is computed to be 
\be
 H 
&=&\f{1}{4\pi\alpha'}\int_0^\ell d\sigma\Bigg[(2\pi\alpha')^2G^{MN}\Pi_M\Pi_N+2(2\pi\alpha')G^{MN}\Pi_MS_N+2(2\pi\alpha')G^{MN}\Pi_MB_{NP}\p_\sigma X^P\non\\
&&+2G^{MN}B_{MP}\p_\sigma X^PS_N+G^{MN}B_{MP}B_{NQ}\p_\sigma X^P\p_\sigma X^Q+G^{MN}S_MS_N+G_{MN}\p_\sigma X^M\p_\sigma X^N\non\\
&&+iG_{MN}\psi^N_+\p_\sigma\psi^M_++2T_P\p_\sigma X^P-ig_{AB}\lambda_-^A\p_\sigma\lambda^B_--\frac{1}{2}F_{MN;AB}\psi_+^M\psi_+^N\lambda_-^A\lambda_-^B\Bigg]\label{85}
\ee
where,
\be
T_M\,\equiv\, \f{i}{2}G_{PQ}\tilde\Gamma^P_{-NM}\psi_+^Q\psi_+^N-\f{i}{2}g_{AB}(\hat A_M)^B_{\;\;C}\lambda_-^A\lambda_-^C\label{TMhet}
\ee
For doing calculations, it is useful to note that the terms in the Hamiltonian \eqref{85} which do not involve $\lambda_-^A$ are same as the corresponding terms in the type II Hamiltonian \eqref{66}.

\vspace*{.07in}We can now simplify the interaction terms for the gauge fields $A_{\mu m},B_{\mu m}$ and $(\hat A_M)^B_{\;\;C}$. We again organise them in two kind of terms. One in which strings interact with one external gauge field and the other in which they interact with their field strength. They are given by
\be
H_1&=& \f{1}{2\pi\alpha'}\int_0^\ell d\sigma \biggl[A_{\mu i}\Bigl\{-(2\pi\alpha')^2\Pi^\mu\Pi^i-(2\pi\alpha')\Pi^\mu B^i_{\;n}\p_\sigma X^n+\p_\sigma X^i \p_\sigma X^\mu\non\\
&&+\f{i}{2}\Bigl(  \psi^\mu_+\p_\sigma\psi^i_+ + \psi^i_+\p_\sigma\psi^\mu_+\Bigl) \Bigl\}\ +\ (\hat A_\mu)_{CD}\Bigl\{ -\f{i}{2}(2\pi\alpha')\Pi^\mu\lambda_-^{CD} -\f{i}{2}\p_\sigma X^\mu\lambda_-^{CD}  \Bigl\}\non\\
&&+B_{\mu i}\Bigl\{  (2\pi\alpha') \Pi^\mu \p_\sigma X^i+(2\pi\alpha')\Pi^i\p_\sigma X^\mu  +g^{ij}B_{jk}\p_\sigma X^k \p_\sigma  X^\mu\Bigl\}\biggl]\label{4.5.113edhd}
\ee
and,
\be
H_2
&=&- \f{i}{4}\int_0^\ell d\sigma \biggl[F^A_{\mu\nu; i}\Bigl\{\Pi^\mu\psi_+^{\nu i}-\Pi^i\psi_+^{\mu \nu}-\f{1}{2\pi\alpha'}\Bigl( B^i_{\;\;\;j}\p_\sigma X^j\psi_+^{\mu\nu}+\p_\sigma X^\mu\psi_+^{\nu i} -\p_\sigma X^i\psi_+^{\mu \nu} \Bigl)\Bigl\}\non\\
&&+F^B_{\mu\nu; i}\Bigl\{ -2\Pi^\mu \psi_+^{\nu i}+\Pi^m\psi_+^{\mu\nu}  +\f{1}{2\pi\alpha'}\Bigl(B^i_{\;\;\;j}\p_\sigma X^j\psi_+^{\mu\nu}+2\p_\sigma X^\mu\psi_+^{\nu i} -\p_\sigma X^i\psi_+^{\mu \nu}\Bigl) \Bigl\}\non\\
&&-\f{i}{2\pi\alpha'}F_{\mu\nu;CD}\psi_+^{\mu\nu}\lambda_-^{CD}\biggl]
\ee
We only focus on the gauge fields $A_{\mu i}$ and $B_{\mu i}$ and work with the rescaled Hamiltonian introduced in equation \eqref{32}. The expectation value of ${\cal H}_1$ between the two generic string states is again exactly identical to the calculation of the bosonic case since the term involving the $\psi_+^\mu$ and $\lambda_-^A$ fields do not contribute. Thus, we need only to focus on the expectation value of ${\cal H}_2$. We first consider the case $B_{\mu i}=0$. In this case, by looking at the expressions, we find that the calculation will be exactly identical to the type II case except that we need to drop the terms involving $\psi_-^M$. This gives
\be
\langle \phi|{\cal H}_2^A|\phi\rangle
&=& \f{i \ell}{2(2\pi\alpha')^2m}F^A_{\mu\nu; i}\int_0^\ell d\sigma\bigl\langle \phi\bigl|\p_-X^i\psi_+^{\mu}\psi_+^\nu \bigl|\phi\bigl\rangle\non\\
&=& \frac{i}{2m}F^A_{\mu\nu; i}\sum_{r\in \mathbb Z+a}\bigl\langle \phi\bigl|p^i_R\bar \psi^\mu_r\bar\psi^\nu_{-r}\bigl|\phi\bigl\rangle\non\\
&=& -\f{1}{2\,m}F^A_{\mu\nu; i}\bigl\langle \phi\bigl|p^i_RK_L^{\mu\nu}\bigl|\phi\bigl\rangle\label{91}
\ee
where $K_L$ is defined in equation \eqref{75}.

\vspace*{.07in}Similarly, the calculation for non zero $B_{\mu i}$ proceeds in the same way as type II case except that we need to forget about the terms involving $\psi_-^\mu$. This gives 
\be
\langle \phi|{\cal H}_2^B|\phi\rangle
&=& -\f{1}{2 m}F_{\mu\nu; i}^B\bigl\langle \phi\bigl|p^i_RK_L^{\mu\nu}\bigl|\phi\bigl\rangle\label{92}
\ee
One point to note about the above results is that $\lambda_-^A$ are Lorentz singlets. Hence, they do not contribute to the spin angular momentum which is reflected in the above expressions.

Finally by combining equations \eqref{91} and \eqref{92} with the contribution coming from $H_1$ as given in equations \eqref{hia} and \eqref{hib} respectively, we find that these expressions, i.e. \eqref{hia} and \eqref{hib},  
are still valid in Heterotic sigma model but with the spin operators replaced by the appropriate expressions.  

\section{String Amplitudes}\label{sec:stringAmpl}

Gyromagnetic factors can be extracted also from string 3pt amplitudes. It turns out that the value of the gyromagnetic factor is entirely encoded within the graviton vertex operator which upon dimensional reduction produces a contribution to the effective string action precisely of the type \eqref{generalGyro}.

The fact that gyromagnetic factors are entirely encoded in the graviton vertex operator show that consistent electromagnetic couplings are related to consistent minimal couplings to gravity before compactification. Uniqueness of the minimal coupling to gravity than translates into highly constrained gyromagnetic factors.

Therefore, in this section we compute the gyromagnetic ratios of arbitrary higher spin states of the bosonic and superstring theories by computing three-point functions of massive high spin states and $U(1)$ gauge fields emerging from the compactification procedure.  The starting point is either the bosonic, superstring or heterotic string theories in the critical dimensions, $d=26$ or $d=10$,  compactified on the torus $T^D\equiv {\cal R}^{D}/2\pi \Lambda_D$, with $\Lambda_D$ the lattice introduced in Sec. \S\ref{bosonic}, to have models with realistic space-time dimensions.

\subsection{Bosonic case}\label{sec:bosonicAmpl}
The compactification generates $U(1)$-gauge fields and we will focus on those coming from the $d$-dimensional gravitons or the Kalb-Ramonds fields  with one index extended along the compact directions and the other one non compact. In bosonic string theory the $26$-dimensional massless fields, graviton, dilaton and Kalb-Ramond,  are described by the vertex operator:
\begin{align}
V_g(z,\bar{z})= 
\varepsilon_{MN}\,\partial_z X_R^M(z)\,\partial_{\bar{z}} X_L^N(\bar{z})e^{\sqrt{\frac{\alpha'}{2}}i\,p_M (X_R^M(z)+X_L^M(\bar{z}))},\label{3.71}
\end{align} 
with $z=e^{2i\frac{\pi}{l} (\tau-\sigma)}$ and $\bar{z}= e^{2i\frac{\pi}{l}(\tau+\sigma)}$. To make explicit the factorization properties of the three point amplitudes in string theory, we decompose, as usual, the polarization of the massless state $\varepsilon_{MN}=\epsilon_M \times \bar{\epsilon}_N$ and we define the left and right vertices as follows:
\begin{align}
V_g(z)&=\epsilon_M \partial_z X_R^M(z)\,e^{\sqrt{\frac{\alpha'}{2}}i\,p_M X_R^M(z)}\,,& V_g(\bar{z})&=\bar{\epsilon}_M \partial_z X_L^M(z)\,e^{\sqrt{\frac{\alpha'}{2}}i\,p_M X_L^M(\bar{z})}.
\end{align}

In the compactification procedure we require that all the components of the $d$-dimensional massless state, remain massless at $d-D$-dimensions. This is achieved by keeping the momentum carried by the vertex different from zero only along the non-compact directions, i.e. $p_M=(p_\mu,\,0)$.  The massive high-spin states, instead, can carry momenta in both compact and non-compact directions\footnote{In the case of the graviton vertex, one could introduce the same momentum notation adopted for the higher spin states,  by defining $p_{R;L}\equiv (\frac{p}{2},\,0)$. In this case the exponential factor of the graviton vertex has to be written in the form $e^{i\sqrt{2\alpha} (p_R\cdot X_R+p_L\cdot X_L)}$. } :
\begin{eqnarray}
p^M_{R,L}\equiv\left(\frac{ p^\mu}{2},\,p_{R,L}^a\right)~~;~~
p_{L,R;a}=\frac{1}{2\sqrt{\alpha'}}\left[n_i+B_{ij}m^j\pm g_{ij} m^j\right]e^{*i}_{a}.
\end{eqnarray}
In equation \eqref{3.71} one can now replace the expression of the field $X(z)$ given by
\be
X_R(z)=\hat{q}-i
\alpha_0\,\ln z +i\sum_{n\neq 0} \frac{\alpha_n}{n} \,z^{-n},
\ee 
with $\alpha_0$ defined in Eq. \eqref{A.1.12} and a similar expression for $X_L(\bar{z})$. Here $\hat{q}$ represents the coordinate of the string center of mass which acts on string states.

The amplitudes that we want to compute involves one gauge field described by the vertex \eqref{3.71} and two identical high-spin states, of the level $N= n_R+n_L$. Here, $n_{L,R}$ are the eigenvalues of the left and right-number operators. The 3-point function can then be expressed in the factorized form
\begin{eqnarray}
A_3=2\kappa_d\left(\frac{2}{\alpha'}\right)\langle n_R,\,p_{1;R}| V_g(z=1) |n_R,\,p_{3;R}\rangle \wedge \langle n_L,\,p_{1;L}| V_g(\bar{z}=1) |n_L,\,p_{3:L}\rangle. \label{1.1}
\end{eqnarray}
Here, we have already used the $SL(2,\mathbb{C})$ invariance of the world-sheet CFT to fix the Koba-Nielsen variables of the massive high-spin vertices to $z=0,\,\infty$, while that of the massless vertex is at the point $z=\bar{z}=1$.

From the above expression \eqref{1.1} we shall now determine the gyromagnetic ratio of the corresponding high-spin states. To this end it will be sufficient to consider an $\alpha'$ expansion of the vertex operator. The leading orders in $\alpha'$ are indeed sufficient to read off the couplings \eqref{APhiPhi}.

Working at the level of the holomorphic part of the above correlator one then gets:
\begin{multline}
\langle n_R,\,p_{1R}| V_g(z=1) |n_R,\,p_{3R}\rangle\\
=\epsilon_M^{(2)}\Bigl\langle n_R,\,p_{1R}\Bigl| :\partial_z X_R^M(1)\Big(1 -\sqrt{\frac{\alpha'}{2}} \sum_{n\neq 0} \frac{p_2\cdot \alpha_n}{n} \Big):\Bigl|n_R,\,p_{3R}+p_{2}\Bigl\rangle\ +\ {\cal O}(\sqrt{\alpha'}).
\end{multline}
Carrying out the algebra substituting the explicit expressions for $\partial_z X_R^M(1)$ one the arrives to:
\begin{multline}
\langle n_R,\,p_{1R}| V_g(z=1) |n_R,\,p_{3R}\rangle\\
= -i\sqrt{\frac{\alpha'}{2}}\Bigl\langle n_R,\,p_{1R}\Bigl|\Big[ \epsilon_M^{(2)} p^M_{3R}- \underbrace{\frac{i}{2} (p_{2M} \epsilon^{(2)}_N -p_{2N}\epsilon^{(2)}_M)}_{F_{MN}(p_2)/2} \hat{S}_R^{M N}\Big]\Bigl|n_R,\,p_{3R}+p_2\Bigl\rangle\ +\ {\cal O}(\sqrt{\alpha'}),
\label{76}
\end{multline}
where we have introduced the right spin operator defined in Eq. \eqref{37} and replaced the momentum operator with its eigenvalue. We have also imposed the on-shell condition $\epsilon^{(2)}\cdot p_2=0$. Similar expressions obviously hold in the left-sector and simply encode the gyromagnetic factors of the open-string states!

From the above open-string correlators one can then recover the associated closed string vertex simply by multiplication:
\begin{align}\label{76b}
	\mathcal{V}_{(\epsilon\bar\epsilon)\Phi\Phi}=-\frac{\alpha^\prime}2\Big\langle\Phi\Big|\left[\epsilon\cdot p_{3;R}- \frac{1}{2}F_{M N}(p_2) \hat{S}_R^{M N}\right]\left[\bar{\epsilon}\cdot p_{3;L}- \frac{1}{2}\bar{F}_{\bar{M} \bar{N}}(p_2) \hat{S}_L^{\bar{M} \bar{N}}\right]\Big|\Phi\Big\rangle\,.
\end{align}
The $(d-D)$-dimensional $U(1)$-gauge fields are obtained by considering polarization tensors with mixed space time indices, one non-compact and the other compact. There exist two possibilities:
\begin{align}
    A_{\mu a}&\equiv \frac{1}{2}\Big( \varepsilon_{\mu}\bar{\varepsilon}_{a}+\varepsilon_{a}\bar{\varepsilon}_{\mu}\Big)\,,\\
    B_{\mu a}&\equiv \frac{1}{2}\Big(\varepsilon_{\mu}\bar{\varepsilon}_{a}-\varepsilon_{a}\bar{\varepsilon}_{\mu}\Big)\,.\label{84}
\end{align}
with $\mu=0\dots d-D-1$ and $a=1,\dots D$.
In terms of these quantities eq.\eqref{76b} becomes:\footnote{Note that since the momentum $p_2$ is entirely non-compact the only possible non-vanishing reduction of $F_{MN}$ is $F_{\mu\nu}$.}
\begin{multline}\label{VAB}    \mathcal{V}_{(A,B)\Phi\Phi}=-\frac{\alpha^\prime}2\Big\langle\Phi\Big|\underbrace{(p_L^a+p_R^a)}_{Q^a}A_a\cdot p_3-\frac{1}{2} F^A_{\mu\nu;
 a} (p_{3L}^aS_R^{\mu\nu} +p_{3R}^a S_L^{\mu\nu} )\\+\underbrace{(p_L^a-p_R^a)}_{\mathcal{Q}^a}B_a\cdot p_3-\frac{1}{2}F^B_{\mu\nu;a} (p_{3L}^a S_R^{\mu\nu} -p_{3R}^aS_L^{\mu\nu} )\Big|\Phi\Big\rangle,
\end{multline}
where we have performed the replacements:
\begin{align}
    \epsilon_\mu\bar{\epsilon}_a&\rightarrow A_{\mu a}+B_{\mu a}\,,& \bar{\epsilon}_\mu{\epsilon}_a&\rightarrow A_{\mu a}-B_{\mu a}\,,
\end{align}
and where we have focused on contribution to the minimal and gyromagnetic couplings so that $F^A_{\mu\nu;a}$, $F^B_{\mu\nu;a}$ are the field strengths in the momentum space  of the gauge fields defined in Eq.\eqref{84}, i.e.:
\begin{align}
F^A_{\mu\nu;a}&=i(p_{2\mu} A_{\nu a}-p_{2\nu} A_{\mu a})\,,& F^B_{\mu\nu;a}&=i(p_{2\mu} B_{\nu a}-p_{2\nu} B_{\mu a})\,.
\end{align}
From eq.~\eqref{VAB} one can read off the charges from the coefficient of the minimal coupling for $A$ and $B$ respectively while the gyromagnetic factor is expressed in terms of the right and left spin-operators therefore producing a particular combination of \eqref{generalGyro} with appropriate coefficients $\alpha^{(j)}$. This will require in particular to relate (depending on the representation considered) $S_R$ and $S_L$ to the canonical spin operators $S_j$.

\subsection{Superstring Case}

The bosonic case case can be easily extended to closed superstring theory. The three-point amplitude to be computed in superstring is the same as in the case of the bosonic theory.  The difference is that the two external high-spin states  are now taken in superghost picture $(-1,-1)$ while the massless states is in the zero picture.
The amplitude where the $SL(2,\mathbb{C})$ invariance has been fixed by choosing $z_1=\infty$, $z_2=1$ and $z_3=0$ is:
\begin{eqnarray}
A_3=2\kappa_d\Big(\frac{2}{\alpha'}\Big)\epsilon_{M}^{(2)}\langle n_R,\,p_{1R}| :\Big[\partial_z X_R^M-\sqrt{\frac{\alpha'}{2}} ip_{2,R}\cdot \psi_-(1)\,\psi_-^M(1)\Big] e^{i\sqrt{\frac{\alpha'}{2}} p_{2}\cdot  X_R(1)}  :  |n_R,\,p_{3R}\rangle \wedge \mbox{L-sect. } \nonumber\\
&&
\end{eqnarray} 
The three point amplitude is now evaluated along the same lines of the bosonic calculation focusing on each open string sector. The expansion of the exponential gives:
\begin{align}
A_3& =2\kappa_d (-i)\epsilon_{M}^{(2)}\langle n_R,\,p_{1R}| :\hat{p}^M-p_{2  N}\Big[ \sum_{n=1}^\infty \frac{\alpha_{-n}^M \alpha_n^N-\alpha_{-n}^N\alpha^M_n}{n} +\delta_{a;0} \frac{1}{2} [\psi_0^M,\,\psi_0^N] \nonumber\\
&\hspace{70pt}+\sum_{r\in \mathbb{Z}+a}( \psi^M_{-r+a}\,\psi^N_{r+a}-\psi^N_{-r+a}\,\psi^M_{r+a})\Big]  :  |n_R,\,p_{3R}+p_{2}\rangle \wedge \mbox{L-sect.}+{\cal O}(\sqrt{\alpha'}) \nonumber\\
&= 2\kappa_d (-i)\epsilon_{M}^{(2)}\langle n_R,\,p_{1R}| \Big[{p}_3^M-i p_{2 N} \hat{S}_R^{M N}\Big]|n_R,\,p_{3R}+p_{2}\rangle\wedge\mbox{L-sect.} +{\cal O}(\sqrt{\alpha'})\,,\label{95}
\end{align} 
with
\begin{align}
\hat{S}_R^{MN}=-i \sum_{n=1}^\infty \frac{\alpha_{-n}^M \alpha_n^N-\alpha_{-n}^N\alpha^M_n}{n} -i \delta_{a;0} \frac{1}{2} [\psi_0^M,\,\psi_0^N] -i\sum_{r=1+a}( \psi^M_{-r+a}\,\psi^N_{r+a}-\psi^N_{-r+a}\,\psi^M_{r+a})\,,
\end{align}
where $a=0,1/2$ in the R and NS sector respectively. 

The right correlator, given in eq.~\eqref{95}, formally coincides with that written in eq.~\eqref{76}, the only difference is in the explicit form of the spin-operators which in the superstring case depends on also fermionic oscillators entering the massive string vertices. Therefore,  the amplitude in superstring, to the leading order in the string slope, is formally identical with the one computed in bosonic string theory. 

Finally,  the vertex operator of the massless state in Heterotic string,  in one sector is equal to the one of the superstring while in the other sector coincides with the one of bosonic theory. The three-point amplitude, in Heterotic string with a massless vertex and two massive high-spin states, to leading order in $\alpha'$, can therefore be obtained by combining the left and right correlators of the bosonic and superstring theories, respectively. This amplitude turns out to be formally identical to the corresponding expressions found in the bosonic and superstring models, the differences are, again, in the explicit realisation of the spin-operators defined on each sector. Consequently, the gyromagnetic factor obtained in the bosonic string case is still valid in these string models. These results, obtained from amplitude calculations, are in agreement with those obtained from the hamiltonian approach developed in the previous sections and confirm the universality of  Eq.s \eqref{hia} and \eqref{hib} from which one can deduce the gyromagnetic ratios of higher-spin particles in string theory.

In the following section we shall consider a few examples with the aim of reducing the combination of $S_R$ and $S_L$ appearing in \eqref{VAB} to the form \eqref{generalGyro} and extract the explicit expressions for the gyromagnetic factors $\alpha^{(j)}$.

\section{Examples}
\label{Examples}

This section is devoted to some explicit examples which involve the states of first Regge trajectory of closed string theory. We shall focus both on totally symmetric fields and hook-fields  and extract the corresponding gyromagnetic ratios from \eqref{VAB}.

\subsection{$p_R=\pm p_L$}

In this special case, where the left and right compact momenta are equal modulo a sign, corresponds to vanishing Kaluza Klein ($p_L=-p_R$) or Winding charges ($p_L=p_R$) defined in equations \eqref{38} and \eqref{50} respectively. In these cases, the expressions of the gyromagnetic ratios simplify considerably. This happens because, in these cases, the relevant interactions depend on the combination $S=S_L+S_R$ which allows us to read off the gyromagnetic ratio for arbitrary elements of the spectrum regardless of the Young Tableaux representation. The gyromagnetic factor in this case is $g=1$. This generalises the results previously obtained in different contexts for massive spin two particles \cite{HIOY:84,Marotta:2019cip,9801072}.

\subsection{Gravitational minimal coupling}
It is interesting to compare the value $g=1$ for the gyromagnetic ratio obtained in the previous subsection with the value $g=2$ which is obtained by requiring consistent HS electromagnetic interactions in constant curvature backgrounds \cite{Cortese:2013lda}. In this section, we shall argue that the universal nature of $g=1$ which is seen in the context of string compactifications is a direct consequence of the uniqueness of gravitational minimal coupling.\footnote{Note that the Kalb-Ramond field can couple to the lowest order in derivatives with two derivative interactions. In the case of totally symmetric fields one has for instance:
\begin{align}
    \mathcal{V}^{(1)}_{B\phi\phi}&=B^{MN}\left(\partial_{M}\phi^\star_{R(s)}\right)\left(\partial_{N}\phi^{R(s)}\right)\,,\\
    \mathcal{V}^{(2)}_{B\phi\phi}&=H^{MNL}\left(\partial_{M}\phi^\star_{NR(s-1)}\right)\left({\phi_L}^{R(s-1)}\right)\,.
\end{align}
While the first coupling gives minimal coupling upon compactification the second contributes to the gyromagnetic ratio. In String theory this coupling is fixed by the double-copy structure of the vertex operator. This is required by T-duality relating the coupling of the Kalb-Ramond field to the coupling of the graviton. Similar considerations can be made for more general representations including fermions. We thank Ashoke Sen for discussions on this point.} This clarifies the universality of $g=1$ in the context of field theory compactifications. Such universality will be lost when considering winding states and more general values of the gyromagnetic factor are possible in these cases as we shall see.

To do so it is useful to analyse which couplings would give a contribution to the gyromagnetic factor upon dimensional reduction. We can restrict to two derivative couplings because these are the only couplings which upon dimensional reduction will produce couplings with a single derivative. Using the classification of cubic couplings obtained in \cite{Boulanger:2008tg,Manvelyan:2010jr,Sagnotti:2010at} we then get three possible two derivative gravitational couplings:
\begin{multline}
    (u_2\cdot u_3)^{s-2}\Big[(u_1\cdot p_{23})^2 (u_2\cdot u_3)^2+\beta_s\,u_1\cdot p_{23} \,(u_2\cdot p_{31}u_3\cdot u_1+u_3\cdot p_{12}u_1\cdot u_2)\\+\gamma_s \,(u_2\cdot p_{31}u_3\cdot u_1+u_3\cdot p_{12}u_1\cdot u_2)^2\Big]\,.
\end{multline}
Their dimensional reduction then gives (focusing on the EM coupling):
\begin{align}
    \frac{2n}{R}(u_2\cdot u_3)^{s-1}\Big[(u_1\cdot p_{23}) (u_2\cdot u_3)+\tfrac{\beta_s}2\, \,(u_2\cdot p_{31}u_3\cdot u_1+u_3\cdot p_{12}u_1\cdot u_2)\Big]\,,
\end{align}
from which one recovers
\begin{align}
    g=\frac{\beta_s}{s}\,.\label{gyroG}
\end{align}
The first observation is that it does not depend on $\gamma_s$ which is the coefficient of the non-minimal coupling proportional to the curvature $R$. The coupling proportional to $\gamma_s$ does not deform the abelian gauge symmetries and is therefore arbitrary in principle. The second observation is that the induced gyromagnetic factor is given by \eqref{gyroG} and one might think that $\beta_s$ could be arbitrary! However the corresponding coupling deforms the gauge transformations of the spin-$s$ field! It must therefore be fixed by the requirement that the induced gauge transformations match the Lie derivative if the HS field are consistently coupled to gravity.

Evaluating the deformation of the gauge transformations using eq. (3.7) of \cite{Joung:2013nma} we obtain:
\be
    &&\hspace*{-.6in}\frac1{s!}u_{3}^{\mu_1}\cdots u_{3}^{\mu_s}(\delta_{\xi}\Phi)_{\mu_1\ldots\mu_s}\non\\
    &=&\frac{1}{s!}(u_2\cdot u_3)^{s-1}\Big[u_1\cdot p_{2} (u_2\cdot u_3)+\tfrac{\beta}2\, (-u_2\cdot p_{1}u_3\cdot u_1+u_3\cdot p_{1}u_1\cdot u_2)\Big]\non\\
    &=&\frac{1}{s!}(u_2\cdot u_3)^{s-1}\Big[{u_1\cdot p_{2} (u_2\cdot u_3)}+{\beta\, u_3\cdot p_{1}u_1\cdot u_2}-\tfrac{\beta}2u_2^{\mu}u_3^\nu(p_{1,\mu}u_{1,\nu}+p_{1,\nu}u_{1,\mu})\Big]
\ee
where the last term that we have set apart can be reabsorbed by a trivial field redefinition being proportional to the symmetrized gradient of the gauge parameter.
Starting instead from the Lie derivative and considering the replacement \eqref{bosegenfunc} we get
\be
    &&\hspace*{-.6in}\frac{1}{s!}u_{3}^{\mu_1}\cdots u_{3}^{\mu_s}\mathcal{L}_{\xi}\phi_{\mu_1\ldots\mu_s}\non\\
    &=&\frac{1}{s!}u_{3}^{\mu_1}\cdots u_{3}^{\mu_s}\left[\xi^\mu\partial_{\mu}\phi_{\mu_1\cdots\mu_s}+s(\partial_{(\mu_1}\xi^{\nu})\phi_{\nu|\mu_2\cdots\mu_s)}\right]\non\\
    &\rightarrow&\frac1{s!^2}u_{3}^{\mu_1}\cdots u_{3}^{\mu_s}(\mathcal{L}_{\xi}\Phi)_{\mu_1\cdots\mu_s}\non\\
    &=&\frac{1}{s!^2}(u_2\cdot u_3)^{s-1}\Big[{u_1\cdot p_{2} (u_2\cdot u_3)}+{s\, u_3\cdot p_{1}u_1\cdot u_2}\Big]
\ee
Requiring that up to trivial redefinition the gauge transformation match among each other forces then $\beta_s=s$. Therefore, we have shown that $g=1$ is a consequence of the uniqueness of gravitational minimal coupling since the minimal coupling is the only coupling which contributes to the gyromagnetic factor upon reduction on the circle. The overall factor in the Lie derivative can be used to fix the overall normalisation of the cubic gravitational coupling.

It is also straightforward to extract $\gamma_s$. We get $\gamma_s=\tfrac{s}2$ in all closed string theories although $\gamma_s$ cannot be fixed by requiring consistency of minimal coupling. It is interesting to note that with this choice of $\gamma_s$ the 2-derivative gravitational coupling of all closed string theories takes the following simple form
\begin{align}
     (u_2\cdot u_3)^{s-2}\Big[u_1\cdot p_{23} \,u_2\cdot u_3+\frac{s}{2} \,\left(u_2\cdot p_{31}u_3\cdot u_1+u_3\cdot p_{12}u_1\cdot u_2\right)\Big]^2\,.
\end{align}
If one considers on the other hand the interaction of open string states with the graviton, one obtains $\gamma_s=0$. It would be interesting to understand if these are the only possible choices for $\gamma_s$ in consistent theories of gravity.

\subsection{Totally symmetric fields}

The case of totally symmetric fields in the first Regge trajectory is the simplest. These are described by Young Tableau diagrams having a single row. In this case, there exist a single gyromagnetic ratio so that both $S_R$ and $S_L$ must contribute to the same structure \eqref{gyroSymm}.

\vspace*{.07in}To extract the gyromagnetic ratio, we can use the following identity
\begin{align}\label{symm_id}
    \Big\langle\Phi\Big|x S^{\mu\nu}_L+y S^{\mu\nu}_R\Big|\Phi\Big\rangle_{\alpha}=\frac1{s}(x \ell_L+y \ell_R)\,\big\langle\Phi\big|S^{\mu\nu}\big|\Phi\big\rangle_u\,,
\end{align}
which is derived in appendix \ref{Young}. This gives
\begin{align}
    \alpha=\frac1{s}(x \ell_L+y \ell_R)\,,
\end{align}
in terms of right and left spins. The gyromagnetic ratio is then obtained by dividing with the charge associated to the corresponding gauge field
\begin{align}
    g^{(a)}_A&=\frac{2}{p_L^a+p_R^a}\frac{p_L^a \ell_R+p_R^a \ell_L}{\ell_R+\ell_L}\,,&  g^{(a)}_B&=\frac{2}{p_L^a-p_R^a}\frac{p_L^a \ell_R-p_R^a \ell_L}{\ell_R+\ell_L}\,.
\end{align}
The above equation shows how for all states which satisfy level matching $p_L^2=p_R^2$, one recovers $g_{A,B}^{(a)}=1$.\footnote{Note that for the gauge field $B$ technically the charge goes to zero but one can still define the gyromagnetic ratio in the limit. It is however possible to obtain values of the gyromagnetic ratio different from one whenever level matching is not satisfied which happens for string states with non-trivial winding along the compact directions.}

\subsection{Mixed-Symmetry}

The mixed symmetry case is more complicated in general but it is the generic case within the string spectrum when compactifications to $d>4$ are considered. We focus here on the example of two row Young Tableaux which appear in the first Regge trajectory of the closed bosonic string.

\vspace*{.07in}In this case, starting from the product of two totally symmetric representations of spin $\ell_R$ and $\ell_L$ with $\ell_R\geq \ell_L$, associated to the first Regge trajectory of the open string, one has to project onto the irreducible component associated to the tableaux $\{\ell_R+\ell_L-k,k\}$.
To obtain the gyromagnetic ratio, we can make use of the identity \footnote{Below, the inner products are of order one in the permutation of indices.}
\begin{align}
    \Big\langle\Phi\Big|x S^{\mu\nu}_L+y S^{\mu\nu}_R\Big|\Phi\Big\rangle=\big\langle\Phi\big|\alpha_1 S_1^{\mu\nu}+a_2S_2^{\mu\nu}\big|\Phi\big\rangle_u\,,
\end{align}
where on the left hand side we have the closed string-correlator and on the right-hand side we used the inner-products among Young Tableaux as described in Appendix \ref{Young}.
Considering the explicit projection on the two-row Young Tableaux we obtain
\be
    \alpha_1=\frac{(\ell_R-k) x+(\ell_L-k) y}{\ell_R+\ell_L-2}\,\qquad,\qquad
    \alpha_2=\frac{(\ell_R-k) y+(\ell_L-k) x}{\ell_R+\ell_L-2}\,.\label{genalph}
\ee
We can then read off the gyromagnetic ratios for the gauge field $A^a_\mu$
\begin{align}
    g_1^{(a)}&=\frac{2}{p_L^a+p_R^a}\frac{(\ell_R-k) p_L^a+(\ell_L-k) p_R^a}{\ell_R+\ell_L-2k}\,,\\
    g_2^{(a)}&=\frac{2}{p_L^a+p_R^a}\frac{(\ell_R-k) p_R^a+(\ell_L-k) p_L^a}{\ell_R+\ell_L-2k}\,,
\end{align}
as well as for the gauge field $B^a_\mu$
\begin{align}
    g_1^{(a)}&=\frac{2}{p_L^a-p_R^a}\frac{(\ell_R-k) p_L^a-(\ell_L-k) p_R^a}{\ell_R+\ell_L-2k}\,,\\
    g_2^{(a)}&=\frac{2}{p_L^a-p_R^a}\frac{-(\ell_R-k) p_R^a+(\ell_L-k) p_L^a}{\ell_R+\ell_L-2k}\,,
\end{align}
Similar expressions follow from for any mixed-symmetry representation in subleading Regge trajectories.

\section{Summary and discussion}\label{s4}
In this work, we considered Bosonic, Type II and Heterotic string theories
compactified on a generic $D$ dimensional torus in the presence of constant moduli. We focused on the interaction between the $U(1)$ gauge fields emerging from the dimensional reduction of the Graviton and Kalb-Ramond fields with the massive HS string states. 
The $d$-dimensional diffeomorphism invariance, broken by the compactification, gives rise to U(1) gauge symmetry to these vector fields. Massive HS string states carrying Kaluza-Klein and Winding numbers are charged with respect to these gauge fields. The cubic couplings between these massive HS fields and the field strengths of the gauge fields is, by definition, proportional to the gyromagnetic ratios.

\vspace*{.07in}We have extracted these couplings in two different ways. We first considered the non-linear sigma-models describing Bosonic, Type II and Heterotic theories in a general compact toroidal background and computed the Hamiltonians giving the interaction between string and the two background gauge fields. We find an expression for the interaction hamiltonian, to leading order in the gauge field and derivatives expansion, which is universal, i.e. same for all the above three string theories, when expressed in terms of the charges and spin operators of the corresponding theories. Expectation value of this hamiltonian, evaluated between two generic physical string states, gives us the gyromagnetic ratios of the corresponding  higher-spin particles. The massive HS states having vanishing KK or Winding charges turn out to have $g=1$ and depend upon the spin and charges in the general case.
For states with mixed symmetry properties described by a generic Young Tableau, there are multiple gyromagnetic couplings, with one per row of the Young diagram. These gyromagnetic couplings were read by projecting onto the appropriate states as described in appendix \ref{Young}.

Our results show how all values of the gyromagnetic factors in closed string theories turn out to be a simple property of the structure of the graviton vertex operator. Interestingly we also point out how to obtain values of $g$ different from one by turning on winding charges. While our approach can be considered top-down, starting from a consistent theory like string theory and deriving the value of the gyromagnetic factor, it would be very interesting to investigate this problem from a bottom-up perspective by deriving how these values of the gyromagnetic factor are consistent with basic principles. For field theory modes this is indeed possible, as one can easily see that $g=1$ is implied by minimal coupling with gravity. It is tempting to think that a similar story should hold also for modes with non-trivial winding, which however cannot be studied in field theory.

It would be interesting to shed some light on the universality of the expression giving the gyromagnetic ratios in other examples such as flux compactifications and string theories on orbifolds breaking partially or totally space-time supersymmetry. A similar analysis can also be performed in principle in curved backgrounds like AdS or dS where one can use the consistency of the boundary correlators to constraint the gyromagnetic factor also in the context of Inflation. Furthermore, it would be interesting to study the connection of these results to fundamental properties like causality directly at the level of the observables focusing e.g. to the AdS and dS cases where causality can be mapped to concrete properties of the dual CFT correlators along similar lines as in \cite{Camanho:2014apa}. We leave this as well as other interesting related questions for future work.

\bigskip

{\bf Acknowledgement:} 
We are thankful to 
Dario Francia, 
and Charlotte Sleight for useful discussions and to Paolo Di Vechia, Rajesh Gopakumar, Massimo Porrati, Ashoke Sen and Kostas Skenderis for key comments on this draft. M.V. is also thankful to the organisers of 15th Kavli Asian Winter School on Strings Particle and Cosmology in which some of the results of this work were presented. The research of M.T. was partially supported by the program  “Rita  Levi  Montalcini”  of the MIUR (Minister for Instruction, University and Research) and the INFN initiative STEFI.

\appendix

\section{Some details of the Hamiltonian Computation}
\label{apps1}

\subsection{ Bosonic sigma model}
\label{bosonicsigma}
The Bosonic sigma model in 26 dimensions in the presence of $G_{MN}$ and $B_{MN}$ is described by
\begin{eqnarray}
S&=&-\frac{1}{4\pi \alpha'} \int d^2\sigma \big[G_{MN}(X)\, \eta^{\alpha\beta} \partial_\alpha X^M\,\partial_\beta X^N+B_{MN}(X) \epsilon^{\alpha\beta} \partial_\alpha X^M\,\partial_\beta X^N\Big]\label{4.1.1i}
\end{eqnarray} 
This action is invariant under a general coordinate transformations and changes by a total derivative under the gauge transformation
\begin{eqnarray}
\delta B_{MN}=\partial_M\Lambda_N-\partial_N\Lambda_M\label{3}
\end{eqnarray}
The equation of motion of $X^M$ coming from the action \eqref{4.1.1i} is
\begin{eqnarray}
\partial_\alpha\partial^\alpha X^P+\Gamma^P_{\;\;MN} \partial_\alpha X^M\,\partial^\alpha X^N- \f{1}{2}H^P_{~MN} \epsilon^{\alpha\beta}\partial_\alpha X^M\,\partial_\beta X^N=0 \label{eq1st}
\end{eqnarray}
with $\Gamma^P_{\;\;MN}$ denoting the Christoffel symbols and
\begin{eqnarray}
H^P_{~MN}=G^{PR}H_{RMN}\qquad;\quad H_{MNP}=\Big(\partial_PB_{MN}+\partial_NB_{PM}+\partial_MB_{NP}\Big)
\end{eqnarray}
The equation of motion \eqref{eq1st} can be easily solved for the compactification described in section \ref{compbasis}. For the background \eqref{7} and at the linear order in the field strengths, only the $ \Gamma^{\mu}_{\nu m}$ components of the Christoffel symbols contribute and are given by
\be
\Gamma^{\mu}_{\nu i}
&=&\f{1}{2}\eta^{\mu \sigma}F^A_{\nu\sigma i}
\ee
Similarly, only the $H_{\mu\nu i}= F^B_{\mu\nu i}$ components of the $H_{MNP}$ contribute. 
With these, the equations of motion for $X^i$ and $X^\mu$ become
\begin{eqnarray}
&&\partial_\alpha\partial^\alpha X^i -\frac{1}{2} g^{ij}F_{\mu\nu;j}^B\epsilon^{\alpha\beta} \partial_\alpha { X}^\mu\partial_\beta { X}^\nu+{\cal O}(F_A^2,F_B^2, F^A F^B)=0\nonumber\\[.2cm]
&&\partial_\alpha \partial^\alpha X^\mu -\eta^{\mu\nu} F^A_{\nu\sigma;i}  \partial_\alpha { X}^\sigma\partial^\alpha { X}^i-\eta^{\mu\nu} F_{\nu\sigma;i}^B\epsilon^{\alpha\beta} \partial_\alpha { X}^\sigma\partial_\beta { X}^i+ {\cal O}(F_A^2,F_B^2,F^A F^B)=0
\end{eqnarray}
These equations can be solved iteratively in the field strengths $F^{A}_{\mu\nu}$ and $F^{B}_{\mu\nu}$. We need the solutions at the zeroth order in the field strengths. However, we note the structure of the solution upto the linear order which can be obtained to be
\begin{eqnarray}
X^i(\uptau,\sigma)&=&X_R^{i}(\sigma^-)+X^{i}_L(\sigma^+)-\frac{1}{2}g^{ij}F_{\mu\nu;j}^B \, X^\mu_R(\sigma^-)\, X^\nu_L(\sigma^+)+{\cal O}(F_A^2,F_B^2,F^A F^B) \nonumber\\
X^\mu(\uptau,\sigma)&=&X_R^{\mu}(\sigma^-)+X^{\mu}_L(\sigma^+)+\frac{1}{2} \eta^{\mu\sigma}F^A_{\sigma \lambda;i} \,{ X}^\lambda(\uptau,\sigma)\,{ X}^i(\uptau,\sigma)\nonumber\\
&&\!+\ \frac{1}{2}\eta^{\mu\sigma}F^B_{\sigma \nu;i} \Bigl( X^\nu_L(\sigma^+)\, X^i_R(\sigma^-)- X^\nu_R(\sigma^-) X^i_L(\sigma^+)\Bigl)
+{\cal O}(F_A^2,F_B^2,F^AF^B)\label{13}
\end{eqnarray}
where $\sigma^\pm=\uptau\pm\sigma$. The combinations $X_R^{i}+X^{i}_L$ and $X_R^{\mu}+X^{\mu}_L$ satisfy the Laplace equation. In this work, we shall not use this more general  solution since  only the leading order terms are required in the expression of the Hamiltonian which gives the interaction of the string with the external gauge fields. By imposing the boundary conditions, the leading order terms can be expressed as 
\begin{eqnarray}
X^i_R(\uptau-\sigma)=\frac{1}{2}x^i+\frac{2\pi\alpha'}{\ell}g^{ij}p_{jR}(\uptau-\sigma)+i\sqrt{\frac{\alpha'}{2}}\sum_{\stackrel{n=-\infty}{n\not=0}}^{\infty}\frac{1}{n}\:\alpha^i_n\:e^{-2i\pi n(\uptau-\sigma) /\ell}\label{pir}
\end{eqnarray}
\vspace*{-.1in}
\begin{eqnarray}
 X^i_L(\uptau+\sigma)=\frac{1}{2}x^i+\frac{2\pi\alpha'}{\ell}g^{ij}p_{jL}(\uptau+\sigma)+i\sqrt{\frac{\alpha'}{2}}\sum_{\stackrel{n=-\infty}{n\not=0}}^{\infty}\frac{1}{n}\:\tilde{\alpha}^i_n\:e^{-2i\pi n(\uptau+\sigma) /\ell}\label{pil}
\end{eqnarray}
The left and right moving momenta in the above expression are given by \cite{9401139,0709.4149}
\be
 p_{iL}=\f{1}{2\sqrt{\alpha'}}\biggl[n_i + (B_{ik}+g_{ik})m^k\biggl]\quad,\quad p_{iR}=\f{1}{2\sqrt{\alpha'}}\biggl[n_i+(B_{ik}-g_{ik})m^k\biggl]\label{16}
 \ee
The $X_R^\mu(\uptau-\sigma)$ and $X_L^\mu(\uptau+\sigma)$ have the same structure as in \eqref{pir} and \eqref{pil} but with $p_{iL}$ and $p_{iR}$ replaced by $p^\mu/2$ where $p^\mu$ denotes the momenta of the state.

\vspace*{.07in}For computing the expectation value of the interacting hamiltonian, following expressions will be useful (denoting $\mathcal{X}^\mu\equiv X_R^\mu+X_L^\mu$)
\be
\p_+X^i_L= \f{\pi}{\ell}\sqrt{2\alpha'} \sum_n\tilde\alpha^i_n\ e^{-2\pi in(\uptau+\sigma)/\ell}\qquad,\quad \p_+{\cal X}^\mu= \f{\pi}{\ell}\sqrt{2\alpha'} \sum_n\tilde\alpha^\mu_n\ e^{-2\pi in(\uptau+\sigma)/\ell}\non\\
\p_-X^i_R= \f{\pi}{\ell}\sqrt{2\alpha'} \sum_n\alpha^i_n\ e^{-2\pi in(\uptau-\sigma)/\ell}\qquad,\quad \p_- {\cal{X}}^\mu= \f{\pi}{\ell}\sqrt{2\alpha'} \sum_n\alpha^\mu_n\ e^{-2\pi in(\uptau-\sigma)/\ell}\label{p+-}
\ee
where, we defined
\be
\tilde\alpha^i_0=\sqrt{2\alpha'} p^i_{L}\qquad,\quad \alpha^i_0=\sqrt{2\alpha'}p_R^i\qquad,\quad \tilde\alpha^\mu_0=\alpha^\mu_0=\sqrt{2\alpha'}\ \f{p^\mu}{2}\label{A.1.12}
\ee
The Virasoro's generators in the compactified theory and to leading order in the gauge fields expansion  are:
\begin{eqnarray}
L_0=\alpha' p_R^2+\frac{\alpha'}{4} p^2+ N-1~~;~~\tilde{L}_0=\alpha'p_L^2+\frac{\alpha'}{4}p^2+ \tilde{N}-1:\nonumber
\end{eqnarray}
where the number operators are given by $N=\sum_{n=1}^\infty \left[g_{\mu\nu}\alpha^\mu_{-n}\,\alpha^\nu_n+g_{ij}\alpha^i_{-n}\,\alpha^j_n\right]$ and with  a similar expression for the left modes of the closed string.The mass-shell condition turns out to be:
\begin{eqnarray}
\frac{\alpha'}{4} M^2= \alpha'g^{ij}\,p_{Ri}\,p_{Rj} +N-1=\alpha'g^{ij}\,p_{Li}\,p_{Lj} +\tilde{N}-1
\end{eqnarray}
which implies the following level matching condition:
\begin{eqnarray}
N-\tilde{N}= n^im_i\label{levelM}
\end{eqnarray}
We counclude this section by observing that the above details do not change the evaluation of the left and right-moving string amplitudes. The only difference is that when considering non-trivial winding one should distinguish left and right momenta and perform the reduction via the following replacements:
\begin{align}
    p_R^M&\to(\frac{1}{2}p^\mu,p_R^i)\,,& p_L^M&\to(\frac{1}{2}p^\mu,p_L^i)\,.
\end{align}
Focusing on the first Regge trajectory one can then use the open string generating function defined in \eqref{OpenBose}, combine two such generating functions, one for left and another for right moving part of the closed string correlator, and impose level matching \eqref{levelM} when expanding the oscillators and obtaining the closed string amplitudes.
\subsection{Type II sigma model}
Superstring theory in a non trivial background is described by the  $(1,1)$-supersymmetric non linear sigma model that realizes  the embedding of the string world-sheet in a space-time with a non trivial  
 metric $G_{MN}$ and an anti-symmetric tensor field $B_{MN}$. The super conformal covariant action describing the supersymmetric sigma model in the superspace notation is given by (see for example Ref. \cite{callan1})
\be
S_{Type II}=\f{1}{4\pi\alpha'}\int d^2\sigma d\theta_+d\theta_-\Bigl[G_{MN}(\Phi)D_+\Phi^MD_-\Phi^N+B_{MN}(\Phi)D_+\Phi^MD_-\Phi^N\Bigl] 
\ee
The world-sheet bosons $X^M$ and the Majorana fermions $\psi^M$ are collected in the superfield 
\begin{eqnarray}
\Phi^M(\sigma,\,\theta)= X^M(\sigma)+\bar{\theta} \psi^M(\sigma) +\frac{1}{2} \bar{\theta}\theta F^M(\sigma)
\end{eqnarray}
where  $F^M(\sigma)$ are auxiliary fields with no dynamics and they will be eliminated by using their equations of motion. The Grassmann coordinates $\theta$ are  two-component Majorana spinors and  $\bar\theta = \theta^t \rho^0$. The components of the spinors are labelled with the indices $A=-,+$. Hence,
\begin{eqnarray}
\psi^M=\left(\begin{array}{l}
\psi^M_-\\
\psi_+^M\end{array}\right)~~;~~\theta=\left(\begin{array}{l}
\theta_-\\
\theta_+\end{array}\right)
\end{eqnarray}
Our convention for the gamma matrices are
\begin{eqnarray}
\rho^0=\left( \begin{array}{cc} 
                   0&-i\\
                   i&0\end{array}\right)~~;~~\rho^1=\left(\begin{array}{cc}
                                                               0&i\\
                                                               i&0\end{array}\right)
\end{eqnarray}
Using these, we find
\begin{eqnarray}
\Phi^M(\sigma,\,\theta)= X^M(\sigma)-i{\theta}_- \psi^M_+(\sigma)+i\theta_+\psi^M_- +i {\theta}_+\theta_- F^M(\sigma)
\end{eqnarray}
The covariant derivative is defined by ($A=-,+$):
\begin{eqnarray}
D_A=\frac{\partial}{\partial \bar{\theta}^A}-i(\rho^\alpha\theta)_A\partial_\alpha\implies
\Bigg\{\begin{array}{l}
D_-=-i\frac {\partial}{\partial \theta_+}-2 \theta_+\partial_-\\
\\
D_+=i\frac{\partial}{\partial \theta_-}+2 \theta_-\partial_+
\end{array}\label{4.3.11}
\end{eqnarray}
Here, we have used the convention
\begin{eqnarray}
\partial_\pm=\frac{1}{2}(\partial_0 \pm \partial_1)~~;~~\partial_0=\partial_++\partial_-~~;~~\partial_1=\partial_+  -\partial_-
\end{eqnarray}
We then have
\begin{eqnarray}
&&D_+\Phi^M= 2\theta_-\partial_+X^M+\psi^M_++\theta_+F^M+2i\theta_-\theta_+\partial_+\psi^M_-\nonumber\\[.3cm]
&&D_-\Phi^M=-2\theta_+\partial_-X^M+\psi^M_-+\theta_-F^M+2i\theta_+\theta_-\partial_-\psi^M_+
\end{eqnarray}
and hence (noting that $\psi_\pm$ are Grassmann valued fields)
\begin{eqnarray}
D_+\Phi^M\,D_-\Phi^N&=&\theta_-\theta_+\Big(-4\partial_+X^M\partial_-X^N -2i\psi_+^M\partial_-\psi_+^N-F^MF^N+2i\partial_+\psi_-^M \psi^N_-\Big)\nonumber\\
&&+\theta_-\Big(2\partial_+X^M\psi_-^N-\psi_+^M\,F^N\Big)+\theta_+\Big(2\partial_-X^N\psi_+^M+\psi_-^NF^M\Big)+\psi_+^M\,\psi_-^N\non
\end{eqnarray}
We can now simplify the two terms in the action. After some manipulation, they are given by 
\begin{eqnarray}
G_{MN}(\Phi) D_+\Phi^M\,D_-\Phi^N
&=&\theta_-\theta_+\biggl[-4G_{MN}\partial_+X^M\partial_-X^N -2iG_{MN}\Bigl(\psi_+^M\nabla_-\psi_+^N+\psi^N_-\nabla_+\psi_-^M \Big)\nonumber\\
&&-G_{MN}F^MF^N-2i G_{QN} \Gamma^Q_{PM}\psi_-^P\,\psi_+^MF^N+\partial_P\partial_QG_{MN}\,\psi_-^P\psi_+^Q\,\psi_+^M\,\psi_-^N\biggl]\non
\end{eqnarray}
and 
\begin{eqnarray}
B_{MN}(\Phi)D_+\Phi^M\,D_-\Phi^N
&=&  \theta_-\theta_+\biggl[B_{MN}\Big(-4\partial_+X^M\partial_-X^N -2i\psi_+^M\partial_-\psi_+^N-2i\psi^N_-\partial_+\psi_-^M \Big)\nonumber\\
&&+2i\partial_P B_{MN} \,\psi_-^P\,\partial_+X^M\psi_-^N+2i\partial_P B_{MN}\psi_+^P\partial_-X^N\psi_+^M\nonumber\\
&&-iH_{MNP}\psi_+^M\,\psi_-^NF^P+\partial_P\partial_QB_{MN} \psi_-^P\psi_+^Q\psi_+^M\,\psi_-^N\biggl]\label{bmnxx}
\end{eqnarray}
where, we defined 
\be
\nabla_-\psi^M_+=\p_-\psi_+^M+\Gamma^M_{PQ}\p_-X^P\psi_+^Q\quad,\qquad 
\nabla_+\psi^M_-=\p_+\psi_-^M+\Gamma^M_{PQ}\p_+X^P\psi_-^Q
\ee

The equation of motion of the auxiliary field $F^M$ gives
\be
F^M= i\Gamma^M_{PQ}\psi^P_+\psi^Q_- \ +\  \f{i}{2}G^{MN}H_{NSR}\psi^R_+\psi^S_- 
\ee
Using the above solution, the terms containing $F^M$, in the action, can be simplified as
\be
&&-G_{MN}F^MF^N-2iG_{MN}\Gamma^M_{RS}\psi_+^R\psi_-^NF^S-iH_{MNP}\psi_+^M\psi_-^NF^P\non\\
&=& \Bigl[G_{MN}\Gamma^M_{PQ}\Gamma^N_{RS}+H_{MPQ}\Gamma^M_{RS}+\f{1}{4}H^T_{\;\;PQ}H_{TRS}\Bigl]\psi_+^Q\psi_+^S\psi_-^P\psi_-^R
\ee
Now, we have 
\begin{eqnarray}
&&\partial_P\partial_QB_{MN} \psi_-^P\psi_+^Q\psi_+^M\psi_-^N +H_{MPQ}\,\Gamma^M_{LS} \psi_+^Q\psi_+^S\psi_-^P\psi_-^L \nonumber\\
&=&\frac{1}{2}\Big[\partial_P\big( \partial_QB_{SL}-\partial_SB_{QL}\big)-\Gamma^M_{PS}H_{QML}-\Gamma^M_{PQ}H_{MSL}\Big]\psi_+^Q\psi_+^S\psi_-^P\psi_-^L\nonumber\\
&=&\frac{1}{2}\nabla_PH_{QSL}\psi_+^Q\psi_+^S\psi_-^P\psi_-^L
\end{eqnarray}
and
\begin{eqnarray}
&&\partial_P\partial_QG_{MN}\psi_-^P\,\psi_+^Q\,\psi_+^M\,\psi_-^N+G_{MN}\Gamma^M_{PQ}\Gamma^N_{RS}\psi_+^Q\psi_+^S\psi_-^P\psi_-^R\non\\
&=&\frac{1}{2} \Bigg[\frac{1}{2}\Big(\partial_L\partial_SG_{MP}-\partial_P\partial_SG_{ML}-\partial_L\partial_MG_{SP}+\partial_P\partial_MG_{SL}
\Big)\non\\
&&+G_{QN}\Big(\Gamma^N_{LS}\Gamma^Q_{PM}- \Gamma^N_{LM}\Gamma^Q_{PS}\Big)\Bigg]\psi_+^S\,\psi_+^M\,\psi_-^L\,\psi_-^P\non\\
&=&-\frac{1}{2} R_{SMLP}\psi_+^S\,\psi_+^M\,\psi_-^L\,\psi_-^P
\end{eqnarray}
We also have 
\be
-2iB_{MN}\psi_+^M\p_-\psi_+^N+2i\p_PB_{MN}\psi_+^P\psi_+^N\p_-X^N=iH_{NPM}\psi_+^P\psi_+^M\p_-X^N-i\p_-(B_{MN}\psi_+^M\psi_+^N)\non\\ \label{bmnvf}
\ee
and,
\be
-2iB_{MN}\psi_-^N\p_+\psi_-^M+2i\p_PB_{MN}\psi_-^P\psi_-^N\p_+X^M=-iH_{NPM}\psi_-^P\psi_-^M\p_+X^N+i\p_+(B_{MN}\psi_-^M\psi_-^N)\non
\ee
Using the above results and performing the grassmann integrals, the action can be written as (ignoring the boundary terms)
\begin{eqnarray}
S&=&\frac{1}{4\pi\alpha'} \int d^2\sigma \bigg[4G_{MN}\partial_+X^M\partial_-X^N+4B_{MN}\partial_+X^M\partial_-X^N +2iG_{MN}\psi_+^M\tilde{\nabla}_-\psi_+^N \nonumber\\
&&\hspace*{.87in}+\ 2iG_{MN} \psi_-^N\tilde{\nabla}_+\psi_-^M+\frac{1}{2}\tilde{R}_{MNPQ}\psi_+^M\psi_+^N\psi_-^P\psi_-^Q\bigg]\label{4.3.45}
\end{eqnarray}
  where, we defined 
\be
\tilde\nabla_-\psi^M_+=\p_-\psi_+^M+\Bigl(\Gamma^M_{PQ}-\f{1}{2}H^M_{\;\;PQ}\Bigl)\psi_+^P\ \p_-X^Q\label{gradpsiplus}\\[.3cm]
\tilde\nabla_+\psi^M_-=\p_+\psi_-^M+\Bigl(\Gamma^M_{PQ}+\f{1}{2}H^M_{\;\;PQ}\Bigl)\psi_-^P\ \p_+X^Q
\ee
and,
\begin{eqnarray}
\tilde R_{MNPQ}=R_{MNPQ}+\f{1}{2} \nabla_PH_{MNQ}-\f{1}{2} \nabla_QH_{MNP}+\frac{1}{4} H_{MRP}H^R_{\;\;QN}-\frac{1}{4} H_{MRQ}H^R_{\;\;PN}
\end{eqnarray}
Upto the quadratic order, the action for the world-sheet field $X^M$ is exactly the same as in the bosonic theory. Hence, upto the orders of our interest, the solutions of the equations of motion for these fields are given by the same expressions as in the closed bosonic case. For the $\psi^M_\pm$ fields, the equations of motion, instead, are given by
\be
\p_\pm\psi_\mp^N+\tilde\Gamma^N_{\pm\;PQ}\psi_\mp^P\p_\pm X^Q-\f{i}{4}\tilde R^N_{\;LSP}\psi_\mp^L\psi_\pm^S\psi_\pm^P=0\label{eompsi1q}
\ee
For our purposes, we need to only solve these at the lowest order, namely $\p_\mp\psi_\pm^M=0$. Moreover, we only need the solution for the non compact directions which are given by
\be
\psi_+^\mu\ =\ \sqrt{\frac{2\pi\alpha'}{\ell}} \sum_{r\in \mathbb Z+a} \bar \psi_r^\mu e^{-2i\pi r(\tau+\sigma)/\ell} \quad,\qquad \psi_-^\mu\ =\ \sqrt{\frac{2\pi\alpha'}{\ell}} \sum_{r\in \mathbb Z+a} \psi_r^\mu e^{-2i\pi r(\tau-\sigma)/\ell}
\ee
where, $a=0$ for R sector and $a=\f{1}{2}$ for the NS sector. 

\vspace*{.07in}Below, we note some results which are useful in computing the Hamiltonian of the compactified theory. In this case also, upto linear order in the fields, the Christoffel symbols which contribute are same as in the bosonic case. Also, only $H_{\mu\nu\, i }= F^B_{\mu\nu,\,i}$ contributes. Hence, we have 
\be
\tilde \Gamma^{\mu}_{\pm\nu i}=\f{1}{2}\eta^{\mu \sigma}F^A_{\nu\sigma,\,i}\pm\f{1}{2}\eta^{\mu\sigma}F^B_{\sigma\nu,\,i }\quad,\qquad\tilde \Gamma^{i}_{+\nu\rho}
=\pm\f{1}{2}g^{ij}F^B_{\nu\rho,\,j}\non
\ee
Next, the $S_M$ defined in \eqref{4.5.tt} are computed to be (using the notation $\psi_\pm^{MN}=\psi_\pm^M\psi_\pm^N$)
\be
S_\mu 
&=&-\f{i}{4}F^A_{\mu\rho,\,i}\Bigl(\psi_+^{\rho i}+\psi_-^{\rho i}\Bigl)-\f{i}{2}F^B_{\rho\mu,\,i}\Bigl(\psi_+^{\rho i}-\psi_-^{\rho i}\Bigl)
\ee
and,
\be
S_i 
&=&-\f{i}{4}F^A_{\mu\rho,\,i}\Bigl(\psi_+^{\rho \mu}+\psi_-^{\rho \mu}\Bigl)+\f{i}{4}F^B_{\rho\mu,\,i }\Bigl(\psi_+^{\rho \mu}-\psi_-^{\rho \mu}\Bigl)
\ee
The $T_M$ defined in \eqref{55tm} have the similar expressions except that we need to change the sign in front of $\psi_+^{MN}$ terms.

\subsection{Heterotic sigma model}
The Heterotic theory has world-sheet supersymmetry in the right moving sector. Consequently, the Sigma model for the Heterotic string theory is described by the action \cite{Sen1, Sen2, Sen3}
\be
S=S_1[\Phi]+S_2[\Phi]+S_3[\Phi,\Lambda_-]
\ee
where,
\be
S_1[\Phi]=\f{2}{4\pi\alpha'}\int d^2\sigma d\theta_-G_{MN}(\Phi)D_+\Phi^M\p_-\Phi^N
\ee
\be
S_2[\Phi]=\f{2}{4\pi\alpha'}\int d^2\sigma d\theta_-B_{MN}(\Phi)D_+\Phi^M\p_-\Phi^N
\ee
\be
S_3[\Phi,\lambda_-]=-\f{i}{4\pi\alpha'}\int d^2\sigma d\theta_-\ g_{AB}(\Phi)\Lambda_-^A(D_++A_+(\Phi))^B_{\;\;\;C}\Lambda_-^C
\ee
and,
\be
\Phi^M(\sigma,\theta_-)=X^M(\sigma)-i\theta_-\psi_+^M\qquad,\qquad D_+=i\f{\p}{\p\theta_-}+2\theta_-\p_+
\ee
\be
\Lambda_-^A(\sigma,\theta_-)=\lambda_-^A+\theta_-f^A(\sigma)\qquad,\qquad A_{\;\;+C}^B=(A_M(\Phi))^B_{\;\;C}D_+\Phi^M
\ee
$\Lambda_-^A$ is anti-commuting. Moreover, only the antisymmetric part of the gauge field $A_M$ contributes. Hence, we can take $(A_M)_{BC}=-(A_M)_{CB}$. We shall now do the component expansion and express the action in terms of the physical component fields. For this, we note that 
\be
D_+\Phi^M =\psi_+^M+2\theta_-\p_+X^M\qquad,\qquad \p_-\Phi^N=\p_-X^N-i\theta_-\p_-\psi_+^N
\ee
\be
G_{MN}(\Phi)&=& G_{MN}(X)-i\theta_-\psi_+^P\p_PG_{MN}\qquad,\qquad B_{MN}(\Phi)= B_{MN}(X)-i\theta_-\psi_+^P\p_PB_{MN}\non\\[.3cm]
g_{AB}(\Phi)&=& g_{AB}(X)-i\theta_-\psi_+^M\p_Mg_{AB}\qquad,\qquad A_M(\Phi)=A_{M}(X)-i\theta_-\psi_+^P\p_PA_{M}\non\\[.3cm]
(A_+)^B_{\;\;C}&=&  (A_M)^B_{\;\;C} \psi_+^M+2\theta_-(A_M)^B_{\;\;C} \p_+X^M-i\theta_-\psi_+^P\psi_+^M(\p_PA_M)^B_{\;\;C}        \non\\[.3cm]
g_{AB}\Lambda_-^A&=&g_{AB}\lambda_-^A+\theta_-g_{AB}f^A-i\theta_-\psi_+^M\lambda_-^A\p_Mg_{AB}\non\\[.3cm]
D_+\Lambda_-^B&=&if^B+2\theta_-\p_+\lambda_-^B
\ee
The three terms in the action can be expressed as
\be
&&\hspace*{-.4in}\f{2}{4\pi\alpha'}G_{MN}(\Phi)D_+\Phi^M\p_-\Phi^N\non\\
&=&\f{2}{4\pi\alpha'}\theta_-\Bigl[2G_{MN}\p_+X^M\p_-X^N+iG_{MN}\psi_+^M\p_-\psi^N_+-i\p_PG_{MN}\psi_+^P\psi_+^M\p_-X^N\Bigl]
\ee
\be
&&\hspace*{-.4in}\f{2}{4\pi\alpha'}B_{MN}(\Phi)D_+\Phi^M\p_-\Phi^N\non\\
&=&\f{2}{4\pi\alpha'}\theta_-\Bigl[2B_{MN}\p_+X^M\p_-X^N+iB_{MN}\psi_+^M\p_-\psi^N_+-i\p_PB_{MN}\psi_+^P\psi_+^M\p_-X^N\Bigl]
\ee
\be
&&\hspace*{-.5in}-\f{i}{4\pi\alpha'}g_{AB}\Lambda_-^A\bigl[D_+\Lambda_-^B+(A_+)^B_{\;\;C}\Lambda_-^C\bigl]\non\\[.2cm]
&=&-\f{i}{4\pi\alpha'}\Bigl[ -2g_{AB}\lambda_-^A\p_+\lambda_-^B  +g_{AB}\lambda_-^A\psi_+^M(A_M)^B_{\;\;C}f^C  -2g_{AB}\lambda_-^A\lambda_-^C(A_M)^B_{\;\;C}\p_+X^M  \non\\
&&\hspace*{.1in}  +ig_{AB}\lambda_-^A\psi_+^P\psi_+^M\lambda_-^C(\p_PA_M)^B_{\;\;C} +ig_{AB}f^Af^B+g_{AB}(A_M)^B_{\;\;C}\psi_+^M\lambda_-^Cf^A\non\\
&&\hspace*{.1in} +\psi_+^M\lambda^A_-\p_Mg_{AB}f^B-i\psi_+^M\lambda^A_-\psi_+^N\lambda_-^C(A_N)^B_{\;\;C}\p_Mg_{AB} \Bigl]
\ee
Now, we have
\be
2ig_{AB}\lambda_-^A\p_+\lambda_-^B  +2ig_{AB}\lambda_-^A\lambda_-^C(A_M)^B_{\;\;C}\p_+X^M
\ =\ 2ig_{AB}\lambda_-^A\hat\nabla_+\lambda_-^B \non
\ee
where,
\be
\hat\nabla_+\lambda_-^B=\p_+\lambda_-^B  +(\hat A_M)^B_{\;\;C}\lambda_-^C\p_+X^M\quad,\qquad (\hat A_M)^B_{\;\;C}=( A_M)^B_{\;\;C}+\f{1}{2}g^{BD}\p_Mg_{DC}
\ee
The terms involving $f^A$ can be simplified as
\be
&&\hspace*{-.5in}  -ig_{AB}\lambda_-^A\psi_+^M(A_M)^B_{\;\;C}f^C +g_{AB}f^Af^B-ig_{AB}(A_M)^B_{\;\;C}\psi_+^M\lambda_-^Cf^A-i\psi_+^M\lambda^A_-\p_Mg_{AB}f^B\non\\[.3cm]
&=& -ig_{AB}f^A\Bigl[if^B+2(\hat A_M)^B_{\;\;C}\psi_+^M\lambda_-^C \Bigl]\label{falambda}
\ee
 The equation of motion of $f^A$ gives 
\be f^A=i(\hat A_M)^A_{\;\;C}\psi_+^M\lambda_-^C
\ee
Substituting this solution in \eqref{falambda} gives
\be
-ig_{AB}f^A\Bigl[if^B+2(\hat A_M)^B_{\;\;C}\psi_+^M\lambda_-^C \Bigl]
&=& -g_{AB}(\hat A_M)^A_{\;\;C}(\hat A_N)^B_{\;\;D}\psi_+^M\psi_+^N\lambda_-^C\lambda_-^D
\label{hgj}
\ee
Using the above results and equation \eqref{bmnvf}, we can express the action as 
\be
S&=&\f{1}{4\pi\alpha'}\int d^2\sigma \Bigl[4G_{MN}\p_+X^M\p_-X^N+4B_{MN}\p_+X^M\p_-X^N+2iG_{MN}\psi_+^M\tilde\nabla_-\psi^N_+  \non\\[.2cm]
&&\hspace*{.5in} +2ig_{AB}\lambda_-^A\hat\nabla_+\lambda_-^B  +\f{1}{2}F_{MN;CD} \psi_+^M\psi_+^N\lambda^C_-\lambda_-^D\Bigl]\label{newacty1}
\ee
where $\tilde\nabla_-\psi^N_+$ is defined in \eqref{gradpsiplus} and 
\be
F_{MN;CD}=\p_M(\hat A_N)_{CD}-\p_N(\hat A_M)_{CD}+(\hat A_M)_{CB}(\hat A_N)^B_{\;\;D}-(\hat A_N)_{CB}(\hat A_M)^B_{\;\;D}
\ee

For the compactification on $T^D$, we again need the mode expansion of the world sheet fields.  At the quadratic order in the world-sheet fields, the equations of motion for the $X^M$ are same as in the bosonic and Type II theories. Hence, upto the orders of our interest, we can use the same expressions for their mode expansion as obtained in these non linear sigma models. The equations of motion of the  $\psi_+^M$, instead, are modified with respect to the corresponding equation in the $(1,1)$ supersymmetric sigma model  by terms containing the background gauge field. However in the following, we shall be only interested in the solution of these equations to the zeroth order in the gauge field $A_M$. Therefore, for $\psi_+^M$ also, we can use the same mode expansion as in the type II theory. Finally, the equation of motion for $\lambda_-^B$ fields are
 \be
\p_+\lambda_-^B+(\hat A_M)^B_{\;\;C}\lambda_-^C\p_+X^M+\f{1}{2}g^{BD}\p_M(g_{CD})\lambda_-^C\p_+X^M-\f{i}{2}g^{BA}F_{MN;AC} \psi_+^M\psi_+^N\lambda^C_-=0
 \ee
This shows that at the lowest order, $\lambda_-^A$ satisfies the same equation as $\psi_-^M$ in type II case. Hence, the solution for these are also given by the same expressions as for $\psi_-^M$. 

\vspace*{.07in}We again use the same compactification ansatz for $G_{MN}$ and $B_{MN}$ as in the bosonic and type II theory. Hence, the expression for the components of $\tilde \Gamma^M_{- NP}$ are same as in the case of type II theory. 
For $S_M$ defined in \eqref{4.5.tth}, we have (using the notation $\psi_+^{MN}=\psi_+^M\psi_+^N$ and $\lambda_-^{AB}=\lambda_-^A\lambda_-^B$)
\be
S_\mu =-\f{i}{4}F_{\mu\rho;i}\psi_+^{\rho i}-\f{i}{2}F^B_{\rho\mu; i}\psi_+^{\rho i}-\f{i}{2}g_{AB}(\hat A_\mu)^B_{\;\;C}\lambda_-^A\lambda_-^C
\ee
and,
\be
S_i 
=-\f{i}{4}F_{\mu\rho;i}\psi_+^{\rho \mu}+\f{i}{4}F^B_{\rho\mu; i}\psi_+^{\rho \mu}-\f{i}{2}g_{AB}(\hat A_i)^B_{\;\;C}\lambda_-^A\lambda_-^C
\ee
Again, the $T_M$ defined in \eqref{TMhet} have the similar expressions except that we need to change the sign in front of $\psi_+^{MN}$ terms. 
\section{ Generalities on 
higher spin states}\label{sec::Intro}

In the following, it will be convenient to use the following spinning polarisations:
\begin{align}
\phi_{\mu_1\ldots\mu_s}(p)=\frac{1}{s!}\,u_{\mu_1}\ldots u_{\mu_s}\ ,\label{bosegenfunc}
\end{align}
in terms of the transverse polarization vector $u^\mu$ depending on the momentum $p$. Upon using point splitting the above replacement allows to transform any spinning correlator into a polynomial in the Lorentz invariant contractions of the vector polarisation $u_i$ associated with each external leg.

Anticipating the results, we shall see that the general pattern of the super-string scattering amplitude generating function can be given as
\begin{align}
\mathcal{A}=\mathcal{B}\,\exp(\mathcal{C})\ ,
\end{align}
where $\exp(\mathcal{C})$ is the same generating function entering the bosonic string case obtained in \cite{Sagnotti:2010at}, while ${\mathcal{B}}$ is overall polynomial in the polarisations whose form appears to be completely specified solely by the fermionic correlators entering the computation in the superstring case.

\vspace*{.07in}It is useful to write down in terms of the polarisations \eqref{bosegenfunc} various cubic couplings which we shall extract from string theory. Focusing on the EM coupling of HS fields which reads:
\begin{align}
   -i e A_\nu \phi_{\mu_1\ldots\mu_s}^*\overleftrightarrow{\partial}^\nu \phi^{\mu_1\ldots\mu_s}&\rightarrow \frac{e}{s!^2} u_1\cdot p_{23}(u_2\cdot u_3)^s\,,
\end{align}
where we remind the reader that $u_3\cdot p_3=0$. Similar expression can be obtained for the non-minimal Pauli coupling:
\begin{align}
    ig \,e s\,F^{\mu\nu}\phi_{\mu\mu_2\ldots\mu_s}^* \phi_{\nu}{}^{\mu_2\ldots\mu_s}&\rightarrow -\frac{e\,g\,s }{(s-1)!^2}(p_1^{\mu}u_1^{\nu}-p_1^{\nu}u_1^{\mu})u_{2,\mu}u_{3,\nu}(u_2\cdot u_3)^{s-1}\\
    &={\frac{e\,g\,s}{2(s-1)!^2}}\,(u_2\cdot p_{31}\,u_1\cdot u_3+u_3\cdot p_{12}\,u_1\cdot u_2)\,(u_2\cdot u_3)^{s-1}\,,\label{B.0.5}
\end{align}
where we introduced the gyromagnetic factor $g$. Similar polynomial expressions can be obtained for any coupling by simply using \eqref{bosegenfunc}.

Here, we  also introduce the representation of the spin-operator on totally symmetric states:
\begin{eqnarray}
(S^{MN})^{M(s)}_{N(s)}
= 2i s \eta^{(M_1[M}\delta^{N]}_{(N_1}\dots \delta^{M_s)}_{N_s)}\label{rspin}
\end{eqnarray}
where $M(s)$ (and $N(s)$) denotes   the completely symmetrized set of indices $\{M_1\dots M_s\}$ and we define
\begin{eqnarray}
u_2^s\cdot S^{MN}\cdot u_3^s\equiv u_{2M_1}\dots u_{2M_s}(S^{MN})^{(M_1\dots M_s)}_{(N_1\dots N_s)}\,u_3^{N_1}\dots u_3^{N_s}=
2i s u_2^{[M}\,u_3^{N]}\, (u_2\cdot u_3)^{s-1}\label{rspin1}
\end{eqnarray}
So far, we have focused on higher spin states in open string theories.  In closed string theories, the polarization of the physical states can be obtained by the factorized product of holomorphic (right) and antiholomorphic (left) sectors,  in the following denoted with $u$ and $\bar{u}$ respectively. Two spin operators, $S_{R;\,L}^{\mu\nu}$ ,  can be introduced in such theories and their action on the polarization of the higher spin states can be equivalently defined as:
\begin{eqnarray}
S_R^{\mu\nu}= i\big(u^\mu\frac{\partial}{\partial u_\nu} -u^\nu\frac{\partial}{\partial u_\mu}\big)~~;\qquad\, S_R^{\mu\nu}= i\big(\bar{u}^\mu\frac{\partial}{\partial \bar{u}_\nu} -\bar{u}^\nu\frac{\partial}{\partial \bar{u}_\mu}\big)
\end{eqnarray}
It is easily seen that\footnote{The notation is $u_{(\mu_1}\dots u_{\mu_l)}=\frac{1}{l!} \sum_{perm. \{1\dots l\}} u_{\mu_1}\dots u_{\mu_l}=u_{\mu_1}\dots u_{\mu_l} $} :
\begin{eqnarray}
S_R^{\mu\nu}\,u_{\mu_1}\dots u_{\mu_l}&=& 2i\sum_{i=1}^l\delta_{\mu_i}^{[\nu}\eta^{\mu] a_i}\delta^{a_1}_{\mu_1}\dots\delta^{a_{i-1}}_{\mu_{i-1}} \delta^{a_{i+1}}_{\mu_{i+1}}\dots \delta^{a_l}_{\mu_l}\,u_{a_1}\dots u_{a_l}\nonumber\\
&=&2i\eta^{a_1[\mu}
\sum_{i=1}^l\delta^{\nu]}_{\mu_i}\frac{1}{(l-1)!}\big( \delta^{a_1}_{(\mu_1}\dots \delta^{a_{i-1}}_{\mu_{i-1}}\,\delta^{a_{i+1}}_{\mu_{i+1}}\dots \delta^{a_l}_{\mu_l)}\big)\,u_{a_1}\dots u_{a_l}\non\\
&=&2i\,l\eta^{a_1[\mu} \delta^{\nu]}_{(\mu_1}\dots  \delta^{a_l}_{\mu_l)}\,u_{a_1}\dots u_{a_l}
\end{eqnarray}
This coincides with Eq. \eqref{rspin}.

The completely symmetric closed string state is defined by the linear combination
\begin{eqnarray}
\phi_{\mu_1\dots \mu_l\nu_1\dots\nu_k}=u_{(\mu_1}\dots u_{\mu_l}\bar{u}_{\nu_1}\dots \bar{u}_{\nu_l)}\label{B.10}
\end{eqnarray}
The action of the spin operators on such states gives:
\begin{eqnarray}
&&S^{\mu\nu}_R\phi_{\mu_1\dots \mu_l\nu_1\dots\nu_k}=
2\,i\,l\,\eta^{a_1[\mu}\delta^{\nu]}_{(\mu_1}\dots \delta^{a_l}_{\mu_l}\delta^{a_{l+1}}_{\nu_1}\dots \delta^{a_{l+k}}_{\nu_k)}\,u_{a_1}\dots u_{a_l}\bar{u}_{a_{l+1}}\dots \bar{u}_{a_{l+k}}\nonumber\\
&&S^{\mu\nu}_L\phi_{\mu_1\dots \mu_l\nu_1\dots\nu_k}=2\,i\,k\,\eta^{a_{l+1}[\mu}\delta^{\nu]}_{(\mu_1}\dots \delta^{a_l}_{\mu_l}\delta^{a_{l+1}}_{\nu_1}\dots \delta^{a_{l+k}}_{\nu_k)}\,u_{a_1}\dots u_{a_l}\bar{u}_{a_{l+1}}\dots \bar{u}_{a_{l+k}}
 \end{eqnarray}
and 
\begin{eqnarray}
(S_L^{\mu\nu}+S_R^{\mu\nu})\phi_{\mu_1\dots \mu_l\nu_1\dots\nu_k}=2\,i\,(l+k)\,\eta^{(a_1[\mu}\delta^{\nu]}_{(\mu_1}\dots \delta^{a_l}_{\mu_l}\delta^{a_{l+1}}_{\nu_1}\dots \delta^{a_{l+k})}_{\nu_k)}\,u_{a_1}\dots u_{a_l}\bar{u}_{a_{l+1}}\dots \bar{u}_{a_{l+k}}\nonumber\\
\label{B.11}
\end{eqnarray}
With the introduction of the auxiliary vector variables $(w,\,\bar{w})$, the totally symmetric state can be represented in the compact form:
\begin{eqnarray}
\phi_{k+l}={\cal N}_{l+k} (u_\cdot w)^k\, (u\cdot \bar{w})^k
\end{eqnarray}
where the left and right polarizations are now identified to get a totally symmetric  tensor. Here ${\cal N}_{l+k}$ is an overall normalization factor that has to be fixed by requiring that the amplitudes give canonically normalized kinetic terms. 

In this representation of spinning particles, the actions of the $S_R$ and $S_L$ spin operators defined in Eq.s \eqref{B.11} are obtained by acting on the states with the operators:
\begin{eqnarray}
S_R^{\mu\nu} = i\Big(w^\mu\frac{\partial}{\partial w_\nu}- w^\nu\frac{\partial}{\partial w_\mu}\Big)~~;~~S_L^{\mu\nu}=i\Big(\bar{w}^\mu\frac{\partial}{\partial \bar{w}_\nu}- \bar{w}^\nu\frac{\partial}{\partial \bar{w}_\mu}\Big)
\end{eqnarray}
introduced in Section \ref{sec:2}.

 \section{Summary of open string amplitudes for the bosonic states in the first Regge trajectory}

In this section we list the open bosonic three point  amplitude in bosonic and super-string  in the non compact and compact background. Closed string amplitudes can be obtained as appropriate squares of the open-string amplitudes.

\subsection{Generating Functions for Vertex operators}

Along the lines of \cite{Sagnotti:2010at} we can use the replacement \eqref{bosegenfunc} to construct \emph{vertex operator generating functions} bosonic string state and NS or R superstring states. Performming the same replacement one can further resum in a single function all vertex operators with different spins.
The bosonic vertex operators after resummation in terms of the vector polarisation \eqref{bosegenfunc} then reads\footnote{In Eq.\eqref{G1},  we have extracted from the string field $X$ the overall  normalization factor $\sqrt{2\alpha'}$ to make such a quantity dimensionless.}
\begin{align}
V(k_R,u)\,=\,\exp\left(i\partial X\cdot u +i\sqrt{2\alpha'}k_R\cdot X\right)\ ,\label{G1}
\end{align}
$k_R$ (and $k_L$ in the closed string)  is the momentum of the string state and the label  $R$  is irrelevant in open-string but in the closed string, where  the vertices are the  product of the right (or holomorphic)  and left (antiholomorphic) sectors, denotes the right-momentum of the particle. In non compact spaces,  $k_R\equiv p$  and $k_R=k_L\equiv p/2$ in open and closed string , respectively. In the compact toroidal background, instead, the momentum long the compact directions is quantized being in the  closed string  equal to $k_{R,\,L} \equiv (p/2, \,p_{R,\,L})$ with $p_{R, \,L}$ defined in Eq. \eqref{16}. 
   
The NS first Regge trajectory generating function in the canonical picture with ghost charge $-1$ can then be given as
\begin{align}
V^{(-1)}(k_R,u)\,=\,:\psi\cdot u\,e^{-\phi}\,\exp\left(i\partial X\cdot u+\,i\sqrt{2\alpha'}k_R\cdot X\right):\ ,\label{G2}
\end{align}
where the dependence on the world-sheet coordinate is understood. 

Similarly, one can construct a generating function of NS vertex operators in the \emph{non-canonical} picture with $0$ ghost charge as
\begin{align}
V^{(0)}(k_R,\xi)=:\!\left\{\left[(\partial\psi\cdot u)+\sqrt{2\alpha^{\prime}}\,k_R\cdot\psi\right]\,(\psi\cdot u)+u\cdot\partial_u\right\}\exp\left( i\partial X\cdot u +\,i\sqrt{2\alpha'}k_R\cdot X\right):\ ,\label{G3}
\end{align}
In the following we will use the vertex operator generating functions so far obtained in order to compute superstring scattering amplitudes. Notice that the transversality condition for the polarization tensors can be translated into a transversality condition for the polarisation $u$ that from now on will be projected on its transverse components. Moreover all vertex generating functions are proportional to the same exponential factor
\begin{align}
\exp\left(i\partial X\cdot u+i\sqrt{2\alpha'}k\cdot X\right)\ ,
\end{align}

Finally it is useful to comment on closed string vertex operators. The key observation is that these can be obtained multiplying together holomorphic and anti-holomorphic open string vertices in the corresponding pictures and imposing level matching:
\begin{align}
    V^{(n,m)}(k_R,\,k_L,u, \bar{u})=\mathfrak{L}\left[V^{(n)}(k_R,u)V^{(m)}(k_L,\bar{u})\right]
\end{align}

An explicit implementation of the level matching operation is given by:
\begin{align}
    \mathfrak{L}\left[f(u)g(\bar{u})\right]=I_{0}\left(2\sqrt{{\lambda}{\bar{\lambda}}}\right)f(\lambda u)g(\bar{\lambda}\bar{u})\Big|_{\lambda=\bar{\lambda}=0}\,.
\end{align}
$I_{0}$ is the Bessel function.  

{When imposing level matching the normalisation of the polarisation tensor is not anymore the one given in \eqref{bosegenfunc}. It is straightforward to fix this normalisation but it is not necessary for the purposes of these notes.

 \subsection{Summary of open string amplitudes for the bosonic states in the first Regge trajectory}

In this section we just list the open bosonic three point  amplitude in bosonic and super-string in  the non compact and compact background. Closed string amplitudes can be obtained as appropriate squares of the open-string amplitudes.

\paragraph{Bosonic Generating Function}
\label{BGF}

The bosonic generating function of three point amplitudes  is obtained by evaluating the string correlator with three vertices defined  in Eq.\eqref{G1}. This contains an explicit dependence on the tree level string Green function. However, three point amplitudes, on-shell, don't have any  dependence  on such quantities and we can write the following expression for the generating function:

\begin{align}\label{OpenBose}
    \mathcal{B}(p_i,u_i)&= e^{u_1\cdot u_2+u_2\cdot u_3+u_3\cdot u_1-\sqrt{\tfrac{\alpha^\prime}{2}}(u_1\cdot p_{23}+u_2\cdot p_{31}+u_3\cdot p_{12})}
\end{align}
with $p_{ij} =p_i-p_j$ and we have used the on-shell condition $p_i\cdot u_i=0$.

\paragraph{Superstring Generating Function}
 The superstring  generating function of three point amplitudes  is instead obtained  by evaluating the correlator with two vertices defined in Eq. \eqref{G2} and one vertex given in Eq. \eqref{G3},   one gets:

\begin{align}
    \mathcal{S}(p_i,u_i)&=\Big(G-u_1\cdot u_2 u_3\cdot u_1-u_2\cdot u_3 u_1\cdot u_2-u_3\cdot u_1 u_2\cdot u_3\Big)\\\nonumber&\hspace{100pt}\times e^{u_1\cdot u_2+u_2\cdot u_3+u_3\cdot u_1-\sqrt{\tfrac{\alpha^\prime}{2}}(u_1\cdot p_{23}+u_2\cdot p_{31}+u_3\cdot p_{12})}
\end{align}
where we have defined the YM combination
\begin{align}
    G=\sqrt{\frac{\alpha^\prime}{2}}\left[u_1\cdot u_2\,u_3\cdot p_{12}+u_2\cdot u_3\,u_1\cdot p_{23}+u_3\cdot u_1\,u_2\cdot p_{31}\right]
\end{align}

Closed string amplitude can be obtained from products of open string amplitudes enforcing level matching. We can obtain both closed bosonic, super and heterotic 3pt amplitudes just implementing level matching when expanding the generating functions.

In order to obtain the dimensional reduction on the  D-dimensional torus we write  the polarization tensor in the form $u\rightarrow (u,\,v)$  being $u$ and $v$ the  non compact and  compact components, respectively.    The amplitude in the compact space is obtained from the corresponding one in the  non compact background by implementing 
dimensional reduction rules. These in open string are: 
\begin{align}
    u_i\cdot u_j&\rightarrow u_i\cdot u_j+v_i\cdot v_j \\
    u_i\cdot p_j&\rightarrow u_i\cdot p_{j}+ v_i\cdot  p_{j} \label{R1}
\end{align}
plus cyclic, where  $p_{j}= \frac{n_j}{R}$ being $n_j$  the KK level and $R$ the compactification length. The corresponding rules in closed-string are defined in Eq. \eqref{redrules}.

In the following we shall extract the part of the coupling which involves one derivative and compare it with \eqref{B.0.5} extracting the corresponding gyromagnetic factor $g$. This can be obtained by taylor expanding the generating functions presented in this section. This is straightforward using the series expansion of exponential.
\paragraph{Open bosonic string}

 The scattering amplitude with one massless state  interacting with two higher-spin states of the leading Regge-trajectory  is obtained by expanding the exponential in Eq. \eqref{OpenBose} and keeping the terms linear in photon polarization and of order $u_{1;2}^s$. The expression al leading order in the string slope turns out to be:  

\begin{eqnarray}
   &&\frac1{s!} (u_2\cdot u_3)^{s-1}\sqrt{\frac{\alpha'}{2}} \left[u_1\cdot p_{23} u_2\cdot u_3+s(u_2\cdot p_{31}u_3\cdot u_1+u_3\cdot p_{12}u_1\cdot u_2)\right]\nonumber\\
   && \frac1{s!} (u_2\cdot u_3)^{s-1}\sqrt{\frac{\alpha'}{2}} \left[u_1\cdot p_{23} u_2\cdot u_3 -  s(u_{1M}\,p_{1N}-u_{1N}\,p_{1M})u_3^{[M}\,u_2^{N]}\right]\label{3open}
\end{eqnarray}
  Here,  $u_1$ is the photon polarization while $u_{2,3}$ are the polarizations of the massive particles.
Eq.\eqref{rspin1}  allows to write  Eq.\eqref{3open} as follows:
\begin{eqnarray}
\frac{1}{s!} \sqrt{\frac{\alpha'}{2}} \left[u_1\cdot p_{23} (u_2\cdot u_3)^s +\frac{i}{2}(u_{1M}\,p_{1N}-u_{1N}\,p_{1M})\,u_3^s\cdot S^{MN}\cdot u_2^s\right]\label{C.1.7}
\end{eqnarray}
which is equivalent to $g=2$. From an effective field theory approach to string amplitudes, the first term of this equation comes from the kinetic term of the higher spin states minimally coupled with an abelian field. The normalization of the string vertices has to be fine-tuned to get canonically normalized kinetic terms. We don't write down explicitly such normalization but it can be straightforward read from Eq. \eqref{C.1.7}. In terms of the physical polarization defined in Eq. \eqref{bosegenfunc} the three-point amplitude takes the form:
\begin{eqnarray}
A_3\sim \Big[ u_1 p_{23} \phi_2\cdot \phi_3 +\frac{1}{2} F_{MN} \phi_{3 a(s)}\, (S^{MN})^{a(s)}_{b(s)} \phi_2^{b(s)} \Big] \label{C.1.8}
\end{eqnarray}
with $F_{MN}$ defined in Eq.\eqref{76}. 

\paragraph{Open superstring string}
Similarly, the superstring amplitude involving one-massless state and two higher spins of the leading Regge trajectory is:

\begin{align}
   \frac1{(s-1)!} (u_2\cdot u_3)^{s-1}\sqrt{\frac{\alpha'}{2}}\left[u_1\cdot p_{23} u_2\cdot u_3+s(u_2\cdot p_{31}u_3\cdot u_1+u_3\cdot p_{12}u_1\cdot u_2)\right]\label{3super}
\end{align}
which is  again and as expected to $g=2$.  This amplitude when written in terms of the polarization of the higher spin states given in Eq. \eqref{bosegenfunc} coincides, at the leading order in the string slope, with Eq. \eqref{C.1.8}.

\paragraph{Closed  strings after reduction on a torus $T^D$}

The closed  amplitudes before the compactification are the same in bosonic, superstring and heterotic string, therefore, being equal to:   
\begin{eqnarray}
A_3^{cl.}\sim&& \left[u_1\cdot k_{R;23} (u_2\cdot u_3)^{s_R} +\frac{i}{2}(u_{1M}\,k_{R;1N}-u_{1N}\,k_{R;1M})\,u_3^{s_R}\cdot S_R^{MN}\cdot u_2^{s_R}\right]\nonumber\\
\times &&\left[\bar{u}_1\cdot k_{L;23} (\bar{u}_2\cdot \bar{u}_3)^{s_L} +\frac{i}{2}(\bar{u}_{1M}\,k_{L;1N}-\bar{u}_{1N}\,k_{L;1M})\,\bar{u}_3^{s_L}\cdot S_L^{MN}\cdot \bar{u}_2^{s_L}\right]
\end{eqnarray}
The compactification  is easily performed by implementing the reduction rules given in Eq.s \eqref{R1}. These correspond to the replacements
\begin{eqnarray}
u_1\cdot k_{r,23}\,(u_2\cdot u_3)^{s_R} \rightarrow&& \Big(\frac{1}{2}u_1\cdot  p_{23} +v_1\cdot p_{R;23}\Big)(u_2\cdot u_3 +v_2\cdot v_3)^{s_R}\nonumber\\
(u_{1M}\,k_{R;1N}-u_{1N}\,k_{R;1M})\,u_3^{s_R}\cdot S_R^{MN}\cdot u_2^{s_R}\rightarrow &&is_R(u_{1\mu}p_{1\nu}-u_{1\nu} p_{1\mu})
u_3^{[\mu}u_2^{\nu]} (u_2\cdot u_3+v_2\cdot v_3)^{s_R-1} \nonumber\\
&&+2 is_R v_{1a} \,p_{1\nu} \,(v_3^a\,u_2^\nu-u_3^\nu \,v_2^a) (u_3 \cdot  u_2 +v_3 \cdot v_2)^{s_R-1}\nonumber\\
&&\label{redrules}
\end{eqnarray}
 Similar relations hold for the antiholomorphic sector.

In the following, we consider in the amplitude only terms relevant for the determination of the  gyromagnetic ratio of the massive fields with respect to the $U(1)$-gauge fields described by the polarization tensors of the form $\varepsilon^{(1)}_{\mu a}\equiv u_{1\,\mu}\,\bar{v}_{1\,a}$ and $\varepsilon^{(1)}_{a\mu}\equiv v_{1\,a}\,\bar{u}_{1\,\mu}$. 
Therefore, we ignore, in the reduction,  terms like the last line of Eq.\eqref{redrules} and those quadratic in the non-compact momenta that correspond to couplings with two derivatives.
 The reduced amplitude turns out to be:
\begin{eqnarray}
&&
A_3^{cl.}\sim \Big[ \frac{1}{2}\Big( p_{23}^\mu\varepsilon_{\mu a} p_{L;23}^a +p_{R;23}^a \varepsilon_{a\mu}p_{23}^\mu\Big)\,(u_2\cdot u_3 +v_2\cdot v_3)^{s_R} (\bar{u}_2\cdot \bar{u}_3+\bar{v}_2\cdot \bar{v}_3)^{s_L}\nonumber\\
&& +\frac{i}{2} s_L(\varepsilon_{a\mu} p_{1\nu}-\varepsilon_{a\nu} p_{1\mu})\,p_{R;23} \,\bar{u}_3^{[\mu}\,\bar{u}_2^{\nu]}\,(u_2\cdot u_3+v_2\cdot v_3)^{s_R}\, (\bar{u}_3 \cdot \bar{u}_2+\bar{v}_2\cdot \bar{v}_3)^{s_L-1}\nonumber\\
&&+\frac{i}{2} s_R(\varepsilon_{\mu a} p_{1\nu}-\varepsilon_{\nu a} p_{1\mu})\,p_{L;23} \,{u}_3^{[\mu}\,{u}_2^{\nu]}\,(u_2\cdot u_3+v_2\cdot v_3)^{s_R-1}\, (\bar{u}_3 \cdot \bar{u}_2+\bar{v}_2\cdot \bar{v}_3)^{s_L}\dots \Big]
\end{eqnarray}
By  introducing the $U(1)$-fields:
\begin{eqnarray}
\varepsilon_{\nu a} =A_{\mu a}+B_{\mu a} ~~;~~\varepsilon_{a \mu}=  A_{\mu a}-B_{\mu a} \end{eqnarray}
with their field strength
\begin{eqnarray}
F^A_{\mu\nu;a}=i(p_\mu\, A_{\nu;a} -p_\nu\, A_{\mu;a})~~;~~F^B_{\mu\nu;a}=i(p_\mu\, B_{\nu;a} -p_\nu\, B_{\mu;a})
\end{eqnarray} 
the amplitudes is rewritten in the form:
\begin{eqnarray}
&&
A_3^{cl.}\sim \Bigg[ (Q^a A_a\cdot p_1+ {\cal Q}^a B_a\cdot p_1)\,(u_2\cdot u_3 +v_2\cdot v_3)^{s_R} (\bar{u}_2\cdot \bar{u}_3+\bar{v}_2\cdot \bar{v}_3)^{s_L} \nonumber\\
&&+\frac{1}{2} F_{\mu\nu;a}^A\Big(p_{L;2}^a \,2i\,s_R u_3^{[\mu}\,u_2^{\nu]}(\bar{u}_3 \cdot \bar{u}_2+\bar{v}_2\cdot \bar{v}_3) + p_{R;2}^a \,2i\,s_L \bar{u}_3^{[\mu}\,\bar{u}_2^{\nu]}\,({u}_3 \cdot {u}_2+{v}_2\cdot {v}_3)\Big)\nonumber\\
&&\times  (u_2\cdot u_3+v_2\cdot v_3)^{s_R-1}\, (\bar{u}_3 \cdot \bar{u}_2+\bar{v}_2\cdot \bar{v}_3)^{s_L-1}\nonumber\\
&&+\frac{1}{2} F_{\mu\nu;a}^B\Big(p_{L;2}^a \,2i\,s_R u_3^{[\mu}\,u_2^{\nu]}(\bar{u}_3 \cdot \bar{u}_2+\bar{v}_2\cdot \bar{v}_3) - p_{R;2}^a \,2i\,s_L \bar{u}_3^{[\mu}\,\bar{u}_2^{\nu]}\,({u}_3 \cdot {u}_2+{v}_2\cdot {v}_3)\Big)\nonumber\\
&&\times  (u_2\cdot u_3+v_2\cdot v_3)^{s_R-1}\, (\bar{u}_3 \cdot \bar{u}_2+\bar{v}_2\cdot \bar{v}_3)^{s_L-1}+\dots \Bigg]
 \end{eqnarray}
 being:
 \begin{eqnarray}
 Q= p_{L;2}+p_{R;2}=-p_{L;3}-p_{R;3}~~;~~{\cal Q}=p_{L;2}-p_{R;2}=-p_{L;3}+p_{R;3}\label{C1}
 \end{eqnarray}
 the charges of the two gauge fields. 
 
The binomial expansion allows to separate in the amplitude  fields with different spin with respect to the reduced $d-D$-Lorenz group\footnote{ $d=10$ or $26$ in superstring and bosonic string, respectively.           $D$ can be taken arbitrary.}, it gives:
\begin{eqnarray}
(u_2\cdot u_3 +v_2\cdot v_3)^{s_R} (\bar{u}_2\cdot \bar{u}_3+\bar{v}_2\cdot \bar{v}_3)^{s_L}= (v_2\cdot v_3)^{s_R}\, (\bar{v}_2\cdot \bar{v}_3)^{s_L}\nonumber\\
+\Big[\sum_{k=0}^{s_L} \sum_{l=1}^{s_R} + \sum_{k=1}^{s_L} \sum_{l=0}^{s_R}\Big] \Big(\begin{array}{c} s_L\\k\end{array}\Big) \Big(\begin{array}{c} s_R\\l\end{array}\Big) 
(\bar{u}_2\cdot \bar{u}_3)^k(u_2\cdot u_3)^l (\bar{v}_2 \cdot \bar{v}_3)^{s_L-k} (v_2\cdot v_3)^{s_R-l}
\end{eqnarray}
We now define.
\begin{eqnarray}
\phi_2^0\,\phi_3^0= \frac{1}{\Gamma(s_R+1)\Gamma(s_L+1)}(v_2\cdot v_3)^{s_R}\, (\bar{v}_2\cdot \bar{v}_3)^{s_L}\label{sp1}
\end{eqnarray}
and
\begin{eqnarray}
\phi_2^{l+k}\cdot\phi_3^{l+k}= \frac{  (\bar{u}_2\cdot \bar{u}_3)^k(u_2\cdot u_3)^l}{k! \,l!} \, \frac{(\bar{v}_2 \cdot \bar{v}_3)^{s_l-k} (v_2\cdot v_3)^{s_R-l}}{(s_L-k)!\,(s_R-l)!} \label{sp2}
\end{eqnarray}
Whit this definition of scalar product we get canonically normalized kinetic terms, being:
\begin{eqnarray}
&&(u_2\cdot u_3 +v_2\cdot v_3)^{s_R} (\bar{u}_2\cdot \bar{u}_3+\bar{v}_2\cdot \bar{v}_3)^{s_L}=\Gamma(s_R+1)\Gamma(s_L+1) \Big[\phi_2^0\,\phi_3^0\nonumber\\
&&+\Big[\sum_{k=0}^{s_L} \sum_{l=1}^{s_R} + \sum_{k=1}^{s_L} \sum_{l=0}^{s_R}\Big]\phi_2^{k+l}\cdot\phi_3^{k+l}\Bigg]
\end{eqnarray} 
In the same way:
\begin{eqnarray}
&&2i\,s_R u_3^{[\mu}\,u_2^{\nu]}(u_2\cdot u_3+v_2\cdot v_3)^{s_R-1} (\bar{u}_3 \cdot \bar{u}_2+\bar{v}_2\cdot \bar{v}_3)^{s_L}= 2i \sum_{l=0}^{ s_R-1} \sum_{k=0}^{ s_L}\frac{s_R\,(s_R-1)!}{(s_R-1-l)! \,l!}\frac{s_L!}{(s_L-k)! \,k!}\nonumber\\
&& \times u_3^{[\mu}\,u_2^{\nu]} (\bar{u}_2\cdot \bar{u}_3)^k(u_2\cdot u_3)^l (\bar{v}_2 \cdot \bar{v}_3)^{s_L-k} (v_2\cdot v_3)^{s_R-1-l}= \sum_{l=1}^{ s_R} \sum_{k=0}^{ s_L}
 \Big(\begin{array}{c} s_L\\k\end{array}\Big) \Big(\begin{array}{c} s_R\\l\end{array}\Big) \nonumber\\
 &&\times  \,2\,i\,l \,u_3^{[\mu}\,u_2^{\nu]}(u_2\cdot u_3)^{l-1} (\bar{u}_2\cdot \bar{u}_3)^k(\bar{v}_2 \cdot \bar{v}_3)^{s_L-k} (v_2\cdot v_3)^{s_R-l}\nonumber\\
 &&= \Gamma(s_R+1)\Gamma(s_L+1)  \sum_{l=1}^{ s_R} \sum_{k=0}^{ s_L} \phi_3^{l+k} \cdot S_R^{\mu\nu}\cdot \phi_2^{l+k}
\end{eqnarray}
Similar relation hold for $S_L^{\mu\nu}$. The amplitude can be written as:
\begin{eqnarray} 
A_3^{cl.}\sim && 
\sum_{l=0}^{ s_R} \sum_{k=0}^{ s_L} \Big[ (Q^a A_a\cdot p_1+ {\cal Q}^a B_a\cdot p_1)\,\phi_2^{l+k}\cdot  \,\phi_3^{l+k}\nonumber\\
&& +\frac{1}{2} F_{\mu\nu;a}^A \Big(p_{L;2}^a \phi^{l+k}_3 S_R^{\mu\nu}\cdot \phi_2^{l+k} +  p_{R;2}^a \phi^{l+k}_3 S_L^{\mu\nu}\cdot \phi_2^{l+k}\Big)\nonumber\\
&&+\frac{1}{2} F_{\mu\nu;a}^B \Big(p_{L;2}^a \phi^{l+k}_3 S_R^{\mu\nu}\cdot \phi_2^{l+k} -  p_{R;2}^a \phi^{l+k}_3 S_L^{\mu\nu}\cdot \phi_2^{l+k}\Big)\dots\Big]
\end{eqnarray}
where we have used the identity:
\begin{eqnarray}
\phi^{0+k}_3 S_R^{\mu\nu}\cdot \phi_2^{0+k}=\phi^{l+0}_3 S_L^{\mu\nu}\cdot \phi_2^{l+0}=0
\end{eqnarray}
Winding and KK-charges are related to the  compact momenta by Eq. \eqref{C1}. It is convenient to rewrite the amplitude only in terms of the charges which are the physical quantities of the theory
\begin{eqnarray}
A_3 \sim && \sum_{l=0}^{ s_R} \sum_{k=0}^{ s_L} \Big\{ (Q^a A_a\cdot p_1+ {\cal Q}^a B_a\cdot p_1)\,\phi_2^{l+k} \cdot\phi_3^{l+k}\nonumber\\
&&+\frac{1}{4} F_{\mu\nu;a}^A \phi^{l+k}_3\cdot  \Big[ Q^a\left(S_R^{\mu\nu}+S_L^{\mu\nu}\right)+{\cal Q}^a\left(S_R^{\mu\nu}-S_L^{\mu\nu}\right)\Big]\cdot \phi_2^{l+k}\nonumber\\
&&+\frac{1}{4} F_{\mu\nu;a}^B  \phi^{l+k}_3 \cdot  \Big[Q^a \left(S_R^{\mu\nu}-S_L^{\mu\nu}\right)+ {\cal Q}^a\left(S^{\mu \nu}_R+S^{\mu\nu}_L\right)\Big]\cdot \phi_2^{l+k}\dots \Big\}
\end{eqnarray}

\section{Young Tableaux and Polynomials}
\label{Young}
In this section, we shall give some details about the mixed-symmetry representations considered in this work. For simplicity, we shall work with projectors and in particular with products of Kronecker-$\delta$'s which we shall label conveniently as
\begin{align}
\delta_{a_1(s_1)\ldots a_n(s_n)}^{b_1(s_1)\ldots b_n(s_n)}\,.
\end{align}
Indices are appropriately projected onto a given mixed-symmetry representation. To each of the above projector one can associate a polynomial built out of auxiliary variables $u_i$ and $w_i$ which play the role of dummy variables associated to each set of totally symmetric indices. For instance the totally symmetric projector reads
\begin{align}
   \mathcal{T}_\ell(u|w)=\mathcal{N}_\ell (u\cdot w)^\ell\,,
\end{align}
where $\mathcal{N}_\ell$ is a normalisation factor which can be fixed by requiring that the above projector squares to itself under contraction of indices, namely
\begin{align}
    \frac{\mathcal{N}_\ell^2}{\ell!}(\partial_{w_1}\cdot\partial_{w_2})^\ell(u_1\cdot w_1)^\ell(u_2\cdot w_2)^\ell=\mathcal{N}_\ell(u_1\cdot u_2)^\ell\,,
\end{align}
implying
\begin{align}
    \mathcal{N}_\ell=\frac1{\ell!}\,.
\end{align}
It is convenient to normalise the contraction of indices for each set of totally symmetric indices as:
\begin{align}
    \prod_i\frac{1}{\ell_i!}(\partial_{w_i}\cdot\partial_{\bar{w}_i})^{\ell_i}\,,
\end{align}
and define the inner product
\begin{align}
    f(w_i)\circ_w g(w_i)=\sum_{\ell_i}\prod_i\frac{1}{\ell_i!}(\partial_{w_i}\cdot\partial_{w_i})^{\ell_i}f(w_i)g(\bar{w}_i)\Big|_{w_i=\bar{w}_i=0}.
\end{align}
Similar projectors can be constructed also for mixed-symmetry fields. For instance the hook projector reads
\begin{align}
    \mathcal{T}_{\ell,1}(u_1,u_2|w_1,w_2)=\mathcal{N}_{\ell,1}(u_1\cdot w_1)^{\ell-1}\left(u_1\cdot w_1\,u_2\cdot w_2-u_1\cdot w_2\,w_2\cdot w_1 \right)\,,
\end{align}
where, the constant $\mathcal{N}_{\ell,1}$ can be fixed to be:
\begin{align}
    \mathcal{N}_{\ell,1}=\frac{1}{\ell+1}\frac{1}{(\ell-1)!}\,,
\end{align}
by requiring that
\begin{align}
    \mathcal{T}_{\ell,1}(u_1,u_2|w_1,w_2)\circ_{w} \mathcal{T}_{\ell,1}(w_1,w_2|v_1,v_2)=\mathcal{T}_{\ell,1}(u_1,u_2|v_1,v_2)\,.
\end{align}
The irreducibility condition is manifestly satisfied and takes the form
\begin{align}
    u_1\cdot\partial_{u_2}\mathcal{T}_{\ell,1}(u_1,u_2|w_1,w_2)=0=w_1\cdot\partial_{w_2}\mathcal{T}_{\ell,1}(u_1,u_2|w_1,w_2)\,.
\end{align}
Sometime, with some abuse of notation, it can be convenient to represent the above projectors using a Dirac notation like:
\begin{align}
    \mathcal{T}_\ell(u_1|w_1)&=\left|u_1\right>_{\ell\ \ell}\left<w_1\right|\,,\\
    \mathcal{T}_{\ell,1}(u_1,u_2|w_1,w_2)&=\left|u_1,u_2\right>_{\ell,1\ \ \ell,1}\left<w_1,w_2\right|\,.
\end{align}
In this work we need to project the tensor product of two totally symmetric representation into irreducible components to extract the corresponding gyromagnetic factors. This is equivalent to decompose into irreducible components the polynomial $(u_1\cdot w_1)^{\ell_1}(u_2\cdot w_2)^{\ell_2}$ and in particular to find the contribution of a given representation to such inner-product. Similar decomposition problems can be worked out more generally but they can be addressed similarly and for this reason will not be considered here.

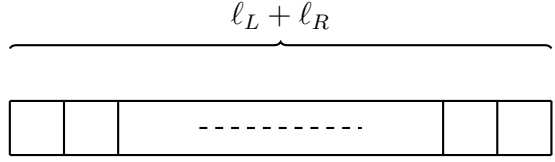
\begin{figure}
\begin{center}\hspace*{-.6in}
\begin{tikzpicture}[scale=.36]
\draw [thick]  (-10,2) -- (10,2);
\draw [thick]  (-10,0) -- (10,0);
\draw [thick]  (-10,2) -- (-10,0);
\draw [thick]  (10,2) -- (10,0);
\draw [thick]  (-8,2) -- (-8,0);
\draw [thick]  (8,2) -- (8,0);
\draw [thick]  (-6,2) -- (-6,0);
\draw [thick]  (6,2) -- (6,0);
\draw [thick,dashed]  (-3,1) -- (3,1);
\draw[decorate, decoration={brace, mirror}, yshift=5ex,thick]  (10,3) -- node[above=0.6ex] {$\ell_L+\ell_R$}  (-10,3);
\end{tikzpicture}
\end{center}
\caption{Young diagram for totally symmetric fields}
\label{Youngsym}
\end{figure}

\vspace*{.07in}To proceed, we need to define intertwiner operators projecting the tensor product of two totally symmetric representations into the possible irreducible representation. Starting from the tensor product of two totally symmetric states, the simplest intertwiner projects onto the totally symmetric component. This gives
\begin{align}
    \mathcal{I}_{\ell_L,\ell_R}^{\ell_L+\ell_R}(u,\bar{u}|w)={\mathcal{N}}_{\ell_R,\ell_L}^{\ell_L+\ell_R}(u\cdot w)^{\ell_L} (\bar{u}\cdot w)^{\ell_R}=(\left|u\right>_{\ell_L}\otimes\left|\bar{u}\right>_{\ell_R})\ {}_{\ell_L+\ell_R}\!\left<w\right|\,,
\end{align}
whose normalization can be obtained by requiring
\begin{align}
    \mathcal{I}_{\ell_L,\ell_R}^{\ell_L+\ell_R}(u_1,u_2|w) \circ_u\mathcal{I}_{\ell_L,\ell_R}^{\ell_L+\ell_R}(u_1,u_2|v)=\mathcal{T}_{\ell_L+\ell_R}(v|w)\,,
\end{align}
which gives
\begin{align}
    {\mathcal{N}}_{\ell_R,\ell_L}^{\ell_L+\ell_R}=\sqrt{\frac1{\ell_1!\ell_2!(\ell_1+\ell_2)!}}\,.
\end{align}
With the above intertwiner operators, it is straightforward to evaluate the string inner product
\begin{align}
    \left\langle\Phi \right|x S_L+y S_R\left|\Phi\right\rangle_{\alpha}\,,
\end{align}
expressing it in terms of the gyromagnetic factors introduced in \S\ref{sec:2}. To do this,
it is sufficient to evaluate the action of $S_{L,R}$ as
\be
    \left\langle\Phi \right|x S^{(u)}_L+y S^{(\bar{u})}_R\left|\Phi\right\rangle_{\alpha}&=& \mathcal{I}_{\ell_L,\ell_R}^{\ell_L+\ell_R}(w|u,\bar{u})\circ_{u}\left[(x S^{(u)}_L+y S_R^{(\bar{u})})\mathcal{I}_{\ell_L,\ell_R}^{\ell_L+\ell_R}(u,\bar{u}|v)\right]\non\\
    &=&\mathcal{T}_{\ell_1+\ell_2}(w|u)\circ_u[\alpha\,S^{(u)}\mathcal{T}_{\ell_1+\ell_2}(u|v)]\non\\
    &=&\left\langle\Phi \right|\alpha\,S\left|\Phi\right\rangle_u\,,
\ee
where the coefficient $\alpha$ can be easily extracted to be
\begin{align}
    \alpha=\frac{x \ell_1+y \ell_2}{\ell_1+\ell_2}\,,
\end{align}
which proves eq. \eqref{symm_id}.

\vspace*{.07in}To obtain the projection of the closed string states into mixed-symmetry components it is again sufficient to derive the corresponding intertwiner operators. In the case of the hook field, one has simply
\begin{align}
    \mathcal{I}_{\ell_L,\ell_R}^{\ell_L+\ell_R-1,1}(u,\bar{u}|w_1,w_2)=\bar{\mathcal{N}}_{\ell_L,\ell_R}^{\ell_L+\ell_R-1,1}(u\cdot w_1)^{\ell_1-1} (\bar{u}\cdot w_1)^{\ell_2-1}\left(u\cdot w_1 \bar{u}\cdot w_2-u\cdot w_2 \bar{u}\cdot w_1\right)\,,
\end{align}
The normalisation can again be obtained by requiring
\begin{align}
    \mathcal{I}_{\ell_L,\ell_R}^{\ell_L+\ell_R-1,1}(u,\bar{u}|w_1,w_2)\circ_u \mathcal{I}_{\ell_L,\ell_R}^{\ell_L+\ell_R-1,1}(u,\bar{u}|v_1,v_2)=\mathcal{T}_{\ell_L+\ell_R-1,1}(w_1,w_2|v_1,v_2)\,,
\end{align}
which gives
\begin{align}
    \mathcal{N}_{\ell_L,\ell_R}^{\ell_L+\ell_R-1,1}=\frac1{\ell_L+\ell_R}\sqrt{\frac{1}{(\ell_L-1)! (\ell_R-1)! (\ell_L+\ell_R-2)!}}\,.
\end{align}

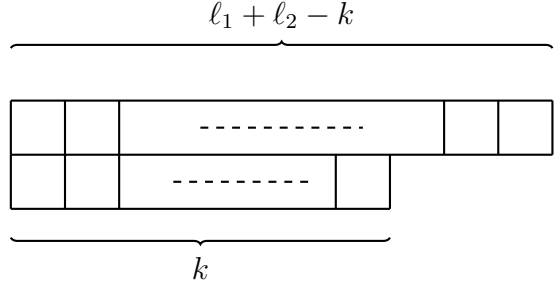
\begin{figure}
\begin{center}\hspace*{-.6in}
\begin{tikzpicture}[scale=.36]
\draw [thick]  (-10,2) -- (10,2);
\draw [thick]  (-10,0) -- (10,0);
\draw [thick]  (-10,2) -- (-10,0);
\draw [thick]  (10,2) -- (10,0);
\draw [thick]  (-8,2) -- (-8,-2);
\draw [thick]  (8,2) -- (8,0);
\draw [thick]  (-6,2) -- (-6,-2);
\draw [thick]  (-10,-2) -- (4,-2);
\draw [thick]  (-10,0) -- (-10,-2);
\draw [thick]  (4,0) -- (4,-2);
\draw [thick]  (2,0) -- (2,-2);

\draw [thick]  (6,2) -- (6,0);
\draw [thick,dashed]  (-3,1) -- (3,1);
\draw [thick,dashed]  (-4,-1) -- (1,-1);

\draw[decorate, decoration={brace, mirror}, yshift=5ex,thick]  (10,3) -- node[above=0.6ex] {$\ell_1+\ell_2-k$}  (-10,3);
\draw[decorate, decoration={brace,mirror}, yshift=5ex,thick]  (-10,-4) -- node[below=0.6ex] {$k$}  (4,-4);

\end{tikzpicture}
\end{center}
\caption{Young diagram with two rows}
\label{Younghook}
\end{figure}

It is now straightforward to evaluate the string-inner product on the intertwiner operator and rewrite it in terms of the canonical spin operators acting on the mixed symmetry representation.
The result reads
\begin{multline}
    \mathcal{I}_{\ell_L,\ell_R}^{(\ell_L+\ell_R-1,1)}(u,\bar{u}|w_1,w_2)\circ_{u}\left[(x S_L^{(u)}+y S_R^{(\bar{u})})\mathcal{I}_{\ell_L,\ell_R}^{(\ell_L+\ell_R-1,1)}(u,\bar{u}|v_1,v_2)\right]\\=\mathcal{T}_{\ell_L+\ell_R-1,1}(u_1,u_2|w_1,w_2)\circ_u\left[(\alpha_1 S_1^{(u_1)}+\alpha_2 S_2^{(u_2)})\mathcal{T}_{\ell_L+\ell_R-1,1}(u_1,u_2|v_1,v_2)\right]\,,
\end{multline}
where in the second line we used $u_1$ and $u_2$ instead of $u$ and $\bar{u}$ to indicate the mixed symmetry dummy variables. The above calculation then gives
\begin{align}
    \alpha_1&=\frac{(\ell_1-1) x+(\ell_2-1) y}{\ell_1+\ell_2-2}\,,\\
    \alpha_2&=\frac{(\ell_1-1) y+(\ell_2-1) x}{\ell_1+\ell_2-2}\,.
\end{align}
These results can be generalised with some effort to the most general case. For example, focusing on the first Regge trajectory of the closed string, we can consider the irreducible projection into an arbitrary two row Yang tableaux of the type $\{\ell_1+\ell_2-k,k\}$. In this generic case we have
\begin{align}
    \mathcal{T}_{\ell,k}=\frac{1}{(\ell-k)! k! (\ell-k+2)_k}(u_1\cdot w_1)^{\ell-k}\left(u_1\cdot w_1\,u_2\cdot w_2-u_1\cdot w_2\,u_2\cdot w_1 \right)^k\,,
\end{align}
together with the intertwiner
\begin{multline}
    \mathcal{I}_{\ell_L,\ell_R}^{\ell_1+\ell_2-k,k}(u,\bar{u}|w_1,w_2)=\frac{1}{k! (\ell_1+\ell_2+2-2k)_k}\sqrt{\frac{1}{(\ell_1-k)! (\ell_2-k)! (\ell_1+\ell_2-2k)!}}\\\times\,(u\cdot w_1)^{\ell_1-k}(\bar{u}\cdot w_1)^{\ell_2-k}\left(u\cdot w_1\,\bar{u}\cdot w_2-u\cdot w_2\,\bar{u}\cdot w_1 \right)^k\,.
\end{multline}
With the above tensor, it is tedious but straightforward to evaluate the inner product between closed string states as
\begin{multline}
    \mathcal{I}_{\ell_L,\ell_R}^{\ell_1+\ell_2-k,k}(u,\bar{u}|w_1,w_2)\circ_u\left[(xS_{L}^{(u)}+yS_R^{(\bar{u})})\mathcal{I}_{\ell_L,\ell_R}^{\ell_1+\ell_2-k,k}(u,\bar{u}|v_1,v_2)\right]\\=\mathcal{T}_{\ell_1+\ell_2-k,k}(w_1,w_2|u_1,u_2)\circ_u\left[(\alpha_1 S_1^{(u_1)}+\alpha_2 S_2^{(u_2)})\mathcal{T}_{\ell_1+\ell_2-k,k}(v_1,v_2|u_1,u_2)\right]\,,
\end{multline}
so that one can find the value of the coefficients $\alpha_{1,2}$ 
\begin{align}
    \alpha_1&=\frac{x (\ell_1-k)+y (\ell_2-k)}{\ell_1+\ell_2-2k}\\
    \alpha_2&=\frac{y (\ell_1-k)+x (\ell_2-k)}{\ell_1+\ell_2-2k}\,,
\end{align}
which is valid for any $k$, $\ell_1$ and $\ell_2$ for which a corresponding representation exists. This is the result given in equation \eqref{genalph}.

\vspace*{.07in} The case of square tableaux needs to be addressed separately or through a limiting procedure. With the above results in hand, we can obtain the gyromagnetic ratios for arbitrary two row representations in the first Regge trajectory of the closed bosonic string. Note that for $k=0,1$ the above results neatly reduce to the special cases discussed above.

\end{document}